\documentclass[12pt]{article}

\usepackage{amssymb,amsmath,amsthm,amsfonts,mathrsfs}
\usepackage{subfigure,epsfig,psfrag,epstopdf}
\usepackage{fullpage,graphicx,color}
\usepackage{algorithm,algorithmic}
\usepackage{tabularx,array,booktabs,bm,multirow}
\usepackage{longtable}
\usepackage{cite,url}
\usepackage[bookmarks=false]{hyperref}
\usepackage[labelfont=bf]{caption}

\newcommand{\mJ}{\mathcal{J}}

\newcommand{\mM}{\mathcal{M}}

\newcommand{\mS}{\mathcal{S}}
\newcommand{\mO}{\mathcal{O}}
\newcommand{\RR}{\mathbb{R}}

\def\bnabla{\boldsymbol{\nabla}}

\def\Dpartial#1#2{\dfrac{\partial #1}{\partial #2}}
\newcommand{\argmin}{\operatorname{argmin}}

\newcommand{\mean}{\operatorname{mean}}

\begin{document}

\title{A Multiscale Optimization Framework for Reconstructing Binary Images using
Multilevel PCA-based Control Space Reduction}
\date{}
\author{{\bf \normalsize Priscilla M.~Koolman} \\ {\it \small College of Engineering \& Science,
Florida Institute of Technology, Melbourne, FL 32901, USA} \and
{\bf \normalsize Vladislav Bukshtynov}\footnote{Corresponding author: \texttt{vbukshtynov@fit.edu}} \\
{\it \small Department of Mathematical Sciences, Florida Institute of Technology, Melbourne, FL 32901, USA}}
\maketitle

\begin{abstract}
An efficient computational approach for optimal reconstructing parameters of
binary-type physical properties for models in biomedical applications
is developed and validated. The methodology includes gradient-based
multiscale optimization with multilevel control space reduction by using
principal component analysis~(PCA) coupled with dynamical control space upscaling.
The reduced dimensional controls are used interchangeably at fine and coarse scales
to accumulate the optimization progress and mitigate side effects at both scales.
Flexibility is achieved through the proposed procedure for calibrating certain
parameters to enhance the performance of the optimization algorithm.
Reduced size of control spaces supplied with adjoint-based gradients obtained
at both scales facilitate the application of this algorithm to models of
higher complexity and also to a broad range of problems in biomedical sciences.
This technique is shown to outperform regular gradient-based methods applied to fine
scale only in terms of both qualities of binary images and computing time.
Performance of the complete computational framework is tested in applications
to 2D inverse problems of cancer detection by the electrical impedance tomography~(EIT).
The results demonstrate the efficient performance of the new method and its
high potential for minimizing possibilities for false positive screening
and improving the overall quality of the EIT-based procedures.

\end{abstract}

\section{Introduction}
\label{sec:intro}

In this work, we propose and validate a computational approach for optimal
reconstructing the physical properties of the media based on any available,
possibly incomplete and noisy, measurements. In particular, this approach
is useful in various applications in biomedical sciences to operate with
physical models characterized by near binary distributions observed for
some of their physical properties, e.g.~heat or electrical conductivity.
The proposed computational framework is using gradient-based multiscale
optimization techniques supplied with multilevel control space reduction
over both fine and coarse scales used interchangeably. Proper ``communication''
established between scales in terms of projecting the solution from one
scale onto another benefits in both quality and computational efficiency
of the obtained results.

As seen in many practical applications, fine scale optimization performed on
fine meshes is able to provide high resolution images for the searched
distributions. Fine meshes also contribute enormously towards increased
sensitivity by enforcing accuracy in computing adjoint states and
constructing adjoint-based gradients if in use. The size of the control
space defined over fine scales may be significantly decreased by applying
any types of parameterization, for instance by using linear transformations
based on available sample solutions (realizations) when applying principal
component analysis~(PCA). However, fine scale
optimization may still suffer from over-parameterization if the problem
is under-determined, i.e.~the number of controls overweighs the size of
available data (measurements). On the other hand, optimization performed
on coarse meshes could arrive at a solution much faster due to the size
of the control space. Usually, the solutions obtained at coarse meshes are
of a low quality and less accurate due to sensitivity naturally ``coarsened''.
In addition to this coarse scale optimization may suffer from being
over-determined if the available data and the size of the control space
are not properly balanced.

Various techniques of multiscale modeling have been used for decades with
proven success for numerous applications in computational mathematics,
engineering and computer science. See
\cite{Steinhauser2017,Weinan2011,Horstemeyer2009}
for a comprehensive description of this general area and an extensive
literature review. There is also a recently growing interest in using
multiscale techniques in applications to biological and biomedical sciences
\cite{Alber2019,Walpole2013,Clancy2016,Tawhai2009}
with limited involvement of methods broadly used in optimization and
control theory.

The proposed multiscale optimization framework utilizes all advantages
mentioned above while using fine and coarse scales. Moreover, using
them both in one process helps us mitigate their side effects.
For example, fine scale solution images may not provide clear
boundaries between regions identified by different physical properties
in space. As a result, a smooth transition cannot provide an accurate
recognition of shapes, e.g.~of cancer-affected regions while solving
an inverse problem of cancer detection~(IPCD). In our computations,
fine scale optimization is used to approximate the location of regions
with high and low values of a physical parameter, namely electrical
conductivity. Projecting solutions onto the coarse scale provides a
dynamical (sharp-edge) filtering to the fine scale images optimized
to better match the available data. The filtered images then projected
back onto the fine scale preserving some information on recent changes
obtained at the coarse scale. In fact, we developed a computationally
efficient procedure for automated scale shifting in order to accumulate
optimally progress obtained at both scales. At the extent of how the
boundaries of the cancerous spots are recovered by projections between
scales with assigned controls at the coarse scale, this approach may be
related to a group of level--set methods which utilize multiscale
techniques and adaptive grids \cite{OsherSethian1988,GibouFedkiwOsher2018,
TsaiOsher2003,ChenChengLinWang2009,Liu2018,Lien2005,Tai2004}.
We also use some notations, main ideas and governing principles of
multiscale parameter estimation (MPE), refer to
\cite{Grimstad2000,Grimstad2003,Cominelli2007,Lien2005}
for some details. In the current paper, we keep the main focus on applying our
new computational approach to IPCD by the Electrical Impedance Tomography~(EIT)
technique, however, the same methodology could be easily applied to a broad range
of problems in biomedical sciences, also in physics, geology, chemistry, etc.

EIT is a rapidly developing non-invasive imaging technique gaining
popularity by enabling various medical applications to perform screening
for cancer detection \cite{Zou2003Review,Brown2003,Adler2008,Holder2004,
Uhlmann2009,Lionheart2004,Abascal2011}.
A well-known fact is that the electrical properties, e.g.~electrical
conductivity or permittivity, of different tissues are changing if the
status of a tissue is changing from healthy to cancer-affected. This
physical phenomenon allows EIT to produce images of biological tissues
by interpreting their response to applied voltages or injected currents
\cite{Brown2003,Cheney1999,Holder2004}. A mathematical model for solving
different types of EIT problems and performing both analytical and
computational analysis of such solutions was suggested in \cite{Cheng1989}.
The inverse EIT problem deals with reconstructing the electrical
conductivity by measuring voltages or currents at electrodes placed on the
surface of a test volume. This so-called Calderon type inverse problem
\cite{Calderon1980} is highly ill-posed, refer to topical review paper
\cite{Borcea2002}. Since 1980s various computational techniques have been
suggested to solve this inverse problem computationally. Recent papers
\cite{Adler2015,Bera2018,Wang2020} review the current state of the art and
the existing open problems associated with EIT and its applications.

This paper proceeds as follows. In Section~\ref{sec:math} we present the
mathematical description of the inverse EIT as an optimization problem
to be solved at both fine and coarse scales by applying control space
reduction using PCA (fine scale) and upscaling via dynamical partitioning
(coarse scale). Procedures for performing multiscale optimization with
interchanging fine and coarse phases are discussed in Section~\ref{sec:OPTframe}.
Model descriptions and detailed computational results including
optimization parameter calibration are presented in Section~\ref{sec:results}.
Concluding remarks are provided in Section~\ref{sec:remarks}.

\section{Mathematical Description}
\label{sec:math}

\subsection{Inverse EIT as an Optimization Problem}
\label{sec:model}

In the recent paper \cite{AbdullaBukshtynovSeif} the inverse EIT problem
is formulated as a PDE-constrained optimization problem in the Besov
spaces framework for which the Fr\'echet gradient and optimality conditions
are derived. The authors also developed the projective gradient method in Besov
spaces and provided extensive numerical analysis for 2D models by implementing
PCA-based solution space re-parameterization and Tikhonov-type regularization.
In our current discussion of the inverse EIT model we use
the same notations as established in \cite{AbdullaBukshtynovSeif}.

Let $\Omega \subset \RR^n, \ n = 2, 3$, be an open and bounded set representing
body and we assume that function $\sigma(x): \, \Omega \rightarrow \RR_+$
represents isotropic electrical conductivity at point $x \in \Omega$. Electrodes
$(E_{\ell})_{\ell=1}^m$ with contact impedances $(Z_{\ell})_{\ell=1}^m
\in \RR^m_+$ are attached to the periphery of the body $\partial \Omega$.
If the so-called ``current--to--voltage'' model is used, electrical currents
$(I_{\ell})_{\ell=1}^m\in \RR^m$ are applied to the electrodes and induce
constant voltages (electrical potentials) $U = (U_{\ell})_{\ell=1}^m \in \RR^m$
on the same electrodes. In this paper, however, we use
the ``voltage--to--current" model where voltages $U_{\ell}$ are applied to
electrodes $E_{\ell}$ to initiate electrical currents $I_{\ell}$.
In either model, it is assumed that both electrical currents and voltages
satisfy the conservation of charge and ground (zero potential) conditions,
respectively
\begin{equation}
  \sum_{\ell=1}^m I_{\ell}=0, \qquad \sum_{\ell=1}^m U_{\ell}=0.
  \label{eq:curr_volt_conds}
\end{equation}

We formulate the inverse EIT (conductivity) problem \cite{Calderon1980}
as a PDE-constrained optimization problem \cite{AbdullaBukshtynovSeif}
by considering minimization of the following cost functional
\begin{equation}
  \mJ(\sigma) = \sum _{\ell=1}^m \left( I_{\ell} - I_{\ell}^* \right)^2,
  \label{eq:cost_functional}
\end{equation}
where $(I_{\ell}^*)_{\ell=1}^m\in \RR^m$ are measurements made for electrical
currents $I_{\ell}$. The latter may be computed as
\begin{equation}
  I_{\ell} = \int_{E_{\ell}}  \sigma(x) \Dpartial{u(x)}{n} \, ds,
  \quad \ell = 1, \ldots, m
  \label{eq:el_current}
\end{equation}
based on conductivity field $\sigma(x)$ set here as a control variable.
A distribution of electrical potential $u(x): \, \Omega \rightarrow \RR$
is obtained as a solution of the elliptic problem
\begin{subequations}
  \begin{alignat}{3}
    \bnabla \cdot \left[ \sigma(x) \bnabla u(x) \right] &= 0,
    && \qquad x \in \Omega \label{eq:forward_1}\\
    \Dpartial{u(x)}{n} &= 0, && \qquad x \in \partial \Omega -
    \bigcup\limits_{\ell=1}^{m} E_{\ell}, \ \ell= 1, \ldots, m
    \label{eq:forward_2}\\
    u(x) +  Z_{\ell} \sigma(x) \Dpartial{u(x)}{n} &= U_{\ell},
    && \qquad x \in E_{\ell}, \ \ell= 1, \ldots, m
    \label{eq:forward_3}
  \end{alignat}
  \label{eq:forward}
\end{subequations}
in which $n$ is an external unit normal vector on $\partial \Omega$. Here
we have to mention a well-known fact that the inverse EIT problem to
identify electrical conductivity $\sigma(x)$ in discretized domain
$\Omega$ with available input data $(I_{\ell}^*)_{\ell=1}^m$ of size $m$
is highly ill-posed. Therefore, we formulate an optimization problem
which is adapted to the situation when the size of input data can be
increased through additional measurements while keeping the size
of the unknown parameters, i.e.~elements in the discretized description
for $\sigma(x)$, fixed. Following the discussion in \cite{AbdullaBukshtynovSeif}
related to the ``rotation scheme'' we set $U^1=U, \ I^1=I$ and consider $m-1$
new permutations of boundary voltages
\begin{equation}
  U^j = (U_j, \ldots, U_{m}, U_1, \ldots, U_{j-1}), \ j = 2, \ldots, m
  \label{eq:permutations}
\end{equation}
applied to electrodes $E_1, E_2, \ldots, E_{m}$ respectively.
Using the ``voltage--to--current'' model allows us to measure
associated currents $I^{j*} = (I^{j*}_1,\ldots,I^{j*}_{m})$.
In addition, the total number of available measurements could be
further increased from $m^2$ up to $K m^2$ by applying
\eqref{eq:permutations} to $K$ different permutations of
potentials within set $U$. Having a new set of $K m$ input data
$(I^{j*})^{Km}_{j=1}$ and in light of the Robin condition
\eqref{eq:forward_3} used together with \eqref{eq:el_current},
we now consider the optimization problem on minimization of
the updated cost functional
\begin{equation}
  \mJ (\sigma) = \sum_{j=1}^{Km} \sum_{\ell=1}^m \beta^j_{\ell}
  \left[ \int_{E_{\ell}} \dfrac{U^j_{\ell}-u^j(x;\sigma)}{Z_{\ell}}
  \, ds - I^{j*}_{\ell} \right]^2,
  \label{eq:cost_functional_final}
\end{equation}
where each function $u^j(\cdot; \sigma), \ j = 1, \ldots, Km$,
solves elliptic PDE problem \eqref{eq:forward_1}--\eqref{eq:forward_3}.
Added weights $\beta^j_{\ell}$ in \eqref{eq:cost_functional_final}
in general allow setting the importance of measurement
$I^{j*}_{\ell}$ (when $\beta^j_{\ell} > 0$), or excluding those
measurements ($\beta^j_{\ell} = 0$) from cost functional $\mJ$
computations. We also could note that the forward EIT problem
\eqref{eq:forward_1}--\eqref{eq:forward_3} together with
\eqref{eq:el_current} may be used to generate various model examples
(synthetic data) for inverse EIT problems to adequately mimic cancer
related diagnoses seen in reality.

Finally, the solution of the optimization problem
\begin{equation}
  \hat \sigma(x) = \underset{\sigma}{\argmin} \ \mJ(\sigma)
  \label{eq:minJ_sigma}
\end{equation}
to minimize cost functional \eqref{eq:cost_functional_final} subject to
PDE constraint \eqref{eq:forward} could be obtained by the first-order
optimality condition which requires the directional differential of
cost functional $\delta \mJ (\sigma; \delta \sigma)$ to vanish for all
perturbations $\delta \sigma$. By invoking the Riesz representation
theorem \cite{Berger1977} in $L_2$ functional space
\begin{equation}
  \delta \mJ (\sigma; \delta \sigma) = \langle \bnabla_{\sigma} \mJ,
  \delta \sigma \rangle_{L_2} = \int_{\Omega} \bnabla_{\sigma} \mJ
  \, \delta \sigma \, d\Omega,
  \label{eq:riesz_L2}
\end{equation}
an iterative algorithm is proposed in \cite{AbdullaBukshtynovSeif}
to solve problem \eqref{eq:minJ_sigma} by means of cost functional
adjoint gradients (with respect to control $\sigma$)
\begin{equation}
  \bnabla_{\sigma} \mJ = - \sum_{j=1}^{Km} \bnabla \psi^j(x)
  \cdot \bnabla u^j(x)
  \label{eq:grad_sigma}
\end{equation}
computed based on solutions $\psi^j(\cdot; \sigma): \, \Omega \rightarrow \RR,
\ j = 1, \ldots, Km$, of the adjoint PDE problem
\begin{subequations}
  \begin{alignat}{3}
    \bnabla \cdot \left[ \sigma(x) \bnabla \psi(x) \right] &= 0,
    && \qquad x \in \Omega \label{eq:adjoint_1}\\
    \Dpartial{\psi(x)}{n} &= 0, && \qquad x \in \partial \Omega -
    \bigcup\limits_{\ell=1}^{m} E_{\ell} \label{eq:adjoint_2}\\
    \psi(x) + Z_{\ell} \Dpartial{\psi(x)}{n} &=
    2 \beta_{\ell} \left[ \int_{E_{\ell}}
    \dfrac{u(x)-U_{\ell}}{Z_{\ell}} \, ds
    + I^*_{\ell} \right], && \qquad
    x \in E_{\ell}, \ \ell= 1, \ldots, m \label{eq:adjoint_3}
  \end{alignat}
  \label{eq:adjoint}
\end{subequations}

\subsection{Fine Scale: PCA-based Control Space Reduction}
\label{sec:fine_mesh}

A well-known problem in numerical optimization is that a spatially
discretized form of the optimization problem discussed in
Section~\ref{sec:model} is over-parameterized even for small size
2D models. To overcome ill-posedness due to over-parameterization
of discretized $\sigma(x)$ along with the fact that the obtained
solutions should also honor any available prior information (such
as available images, etc.), we implement re-parameterization of the
control space based on principal component analysis, also known
as Proper Orthogonal Decomposition (POD) or Karhunen--Lo\`{e}ve (KL)
expansion.

More specifically, PCA enables us to represent control $\sigma(x)$
in terms of uncorrelated variables (components of vector $\xi$)
mapping $\sigma(x)$ and $\xi$ by
\begin{subequations}
  \begin{alignat}{1}
    \sigma &= \Phi \, \xi + \bar \sigma, \label{eq:map_pca_1}\\
    \xi &= \hat \Phi^{-1} (\sigma - \bar \sigma), \label{eq:map_pca_2}
  \end{alignat}
  \label{eq:map_pca}
\end{subequations}
where $\Phi$ is the basis (linear transformation) matrix, $\hat \Phi^{-1}$
is the pseudo-inverse of $\Phi$, and $\bar \sigma$ is the prior mean.
In the PCA used in our numerical experiments, the truncated singular value
decomposition (TSVD) of a (centered) matrix, containing $N_r$ sample
solutions (realizations) $(\sigma^*_n)_{n=1}^{N_r}$, as its columns,
is used to construct the basis matrix $\Phi$. The prior mean is given by
$\bar \sigma = (1/N_r) \sum_{n=1}^{N_r} \sigma^*_n$, see
\cite{AbdullaBukshtynovSeif,Bukshtynov15,JolliffeCadima2016,Jolliffe2002}
for PCA theory in general and details on constructing a complete PCA
representation in particular. The optimization problem initially
defined in Section~\ref{sec:model} is now restated in terms of new model
parameters $\xi \in \RR^{N_{\xi}}$ used in place of control
$\sigma(x)$ as follows
\begin{equation}
  \hat \xi = \underset{\xi}{\argmin} \ \mJ(\xi)
  \label{eq:minJ_xi}
\end{equation}
subject to discretized PDE model \eqref{eq:forward} and using control
mapping \eqref{eq:map_pca} for computing $\mJ(\xi) = \mJ(\sigma(\xi))$.
New gradients $\bnabla_{\xi} \mJ$ of cost
functional $\mJ(\sigma)$ with respect to new control $\xi$ can be
expressed as
\begin{equation}
  \bnabla_{\xi} \mJ = \Phi^T \, \bnabla_{\sigma} \mJ
  \label{eq:grad_xi}
\end{equation}
to define projection of gradients $\bnabla_{\sigma} \mJ$ obtained by
\eqref{eq:grad_sigma} from initial (physical) $\sigma$-space onto the
reduced-dimensional $\xi$-space.

\subsection{Coarse Scale: Control Space Upscaling via Partitioning}
\label{sec:coarse_mesh}

An optimum in employing PCA-based control space reduction on a fine
mesh discussed in Section~\ref{sec:fine_mesh} will be achieved
by finding the minimal (optimal) size of new control $\xi$ to
enable honoring prior information from available sample solutions
\cite{AbdullaBukshtynovSeif,Bukshtynov15}. As often seen in
practical applications, this optimal size cannot prevent the
optimization problem \eqref{eq:minJ_xi} from being still
over-parameterized. Therefore, one will be interested in further
re-parameterization by finding a new control space defined
by a reasonably small number of parameters, and thus, having
fewer local minima.

\begin{figure}[!htb]
  \begin{center}
  \mbox{
  \subfigure[]{\includegraphics[width=0.33\textwidth]{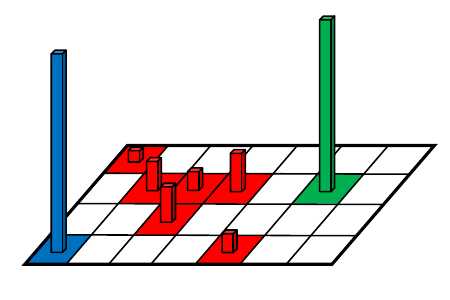}}
  \subfigure[]{\includegraphics[width=0.33\textwidth]{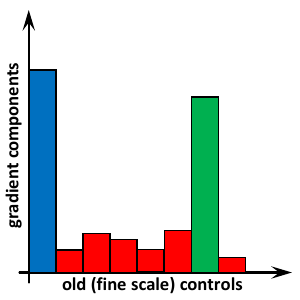}}
  \subfigure[]{\includegraphics[width=0.33\textwidth]{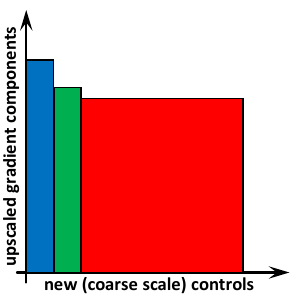}}}
  \end{center}
  \caption{Schematic illustrating the general concept of gradient-based
    multiscale estimation (GBME). Plots in (a,b) show ``noncompetitive''
    vs.~``competitive'' controls identified by associated small (red bars)
    and big (blue and green bars) components of the gradient. A new (cumulative)
    gradient component obtained via control grouping is depicted by the big
    red bar in~(c) with dimensions compounded by summing the respective
    dimensions of small red bars in~(b). For simplicity, all gradients shown
    in (a,b,c) have only positive components.}
  \label{fig:gbmpe}
\end{figure}

As a motivation for our new multiscale approach we used an idea
of gradient-based multiscale (parameter) estimation (GBME) raised
from the general MPE principles \cite{Cominelli2007}.
Figure~\ref{fig:gbmpe} illustrates the
general concept of GBME. It employs various approaches for
gradient-based refinement of the control space for dynamical space
upscaling, i.e.~control grouping, by analysis of changes in the
gradient structure. For example, ``noncompetitive'' controls identified
by relatively small components in the gradient, shown in red in
Figures~\ref{fig:gbmpe}(a,b), are grouped into a new control.
The associated (cumulative) component of the upscaled gradient,
added as a red bar in Figure~\ref{fig:gbmpe}(c), makes the new control
competitive and ``visible'' by other controls shown in blue and green.

Keeping aside for a while a principle by which spatial controls are grouped
(partitioned) in our new approach, let us focus on obtaining upscaled gradients
for new controls $(\zeta_j)_{j=1}^{N_{\zeta}} \in \RR^{N_{\zeta}}_+$. We assume
that control $\sigma(x)$ in problem \eqref{eq:minJ_sigma} is discretized over
the fine mesh containing $N$ elements and represented by a finite set of controls
$(\sigma_i)_{i=1}^N \in \RR^N_+$. This set is then partitioned into $N_{\zeta}$
subsets $C_j, \ j = 1, \ldots, N_{\zeta}$, by selecting (without repetition) $N_j$
controls for $j$-th subset and defining a map, i.e.~fine--to-coarse partition,
\begin{equation}
  \mM: \, (\sigma_i)_{i=1}^N \rightarrow \bigcup \limits_{j=1}^{N_{\zeta}} C_j,
  \quad C_j = \{ \sigma_i : P_{i,j} = 1, \, i = 1, \ldots, N\},
  \quad \sum_{j=1}^{N_{\zeta}} |C_j| = \sum_{j=1}^{N_{\zeta}} N_j = N,
  \label{eq:upscale_map}
\end{equation}
where the partition indicator function is defined as
\begin{equation}
  P_{i,j} = \left\{
  \begin{aligned}
    1, \quad \sigma_i \in C_j,\\
    0, \quad \sigma_i \notin C_j.
  \end{aligned}
  \right.
  \label{eq:part_ind}
\end{equation}

\begin{figure}[!htb]
  \begin{center}
  \mbox{
  \subfigure[]{\includegraphics[width=0.33\textwidth]{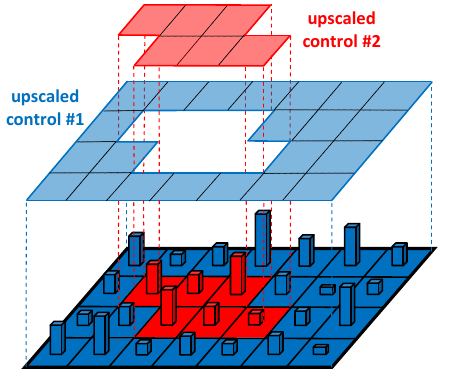}}
  \subfigure[]{\includegraphics[width=0.33\textwidth]{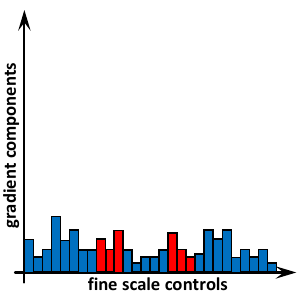}}
  \subfigure[]{\includegraphics[width=0.33\textwidth]{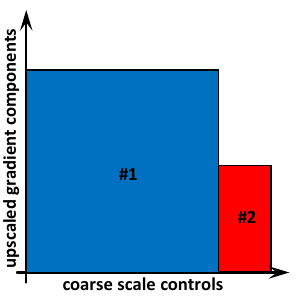}}}
  \end{center}
  \caption{Schematic illustrating the general concept of dynamical control space
    relaxation and obtaining upscaled gradients at coarse scale by making
    2-group assignments for fine scale controls, $N_{\zeta} = 2$. Plots in (a,b)
    show all controls on a fine mesh. Different colors are used to identify controls
    to be included into (blue)~upscaled control~\#1 and (red)~upscaled control~\#2.
    The new (cumulative) gradient components obtained via control grouping
    (partitioning) are depicted by the big blue and red bars in (c)~with
    dimensions compounded by summing the respective dimensions of small blue
    and red bars in (b). For simplicity, all gradients shown in (a,b,c) have
    only positive components.}
  \label{fig:mpe_eit}
\end{figure}

To proceed with gradients, we now consider discretized directional
differential $\delta \tilde \mJ$, obtained from \eqref{eq:riesz_L2}
by the first-order scheme, which is consistent with the discretized form of
domain~$\Omega$ decomposed into $N$ spatial elements
$(\delta \Omega_i)_{i=1}^N$ each of area (or volume in 3D) $\Delta_i$
\begin{equation}
  \delta \mJ (\sigma; \delta \sigma) \approx \delta \tilde \mJ =
  \sum_{i=1}^N \Dpartial{\mJ}{\sigma_i} \Delta_i \, \delta \sigma_i,
  \label{eq:riesz_discr_1}
\end{equation}
where $\delta \sigma_i$ perturbs controls $\sigma_i$. Whenever control
grouping is in place, we assume that all controls $\sigma_i$ within
the same $j$-th group, $\sigma_i \in C_j$, are perturbed equally,
i.e.~$\delta \sigma_i = \delta \zeta_j$ if $P_{i,j} = 1$ for all
$i = 1, \ldots N$ and $j = 1, \ldots N_{\zeta}$. Then one could easily
show that the spatial grouping is fully consistent with the Riesz theorem
\begin{equation}
  \delta \tilde \mJ =
  \sum_{j=1}^{N_{\zeta}} \sum_{i=1}^{N} P_{i,j} \Dpartial{\mJ}{\sigma_i}
  \Delta_i \, \delta \sigma_i =
  \sum_{j=1}^{N_{\zeta}} \Dpartial{\mJ}{\zeta_j} \, \delta \zeta_j =
  \langle \bnabla_{\zeta} \mJ, \delta \zeta \rangle
  \label{eq:riesz_discr_2}
\end{equation}
and upscaled gradients $\bnabla_{\zeta} \mJ$ are computed by summing up
those components of discretized gradients $\bnabla_{\sigma} \mJ$ related
to controls $\sigma_i \in C_j$, i.e.
\begin{equation}
  \Dpartial{\mJ}{\zeta_j} =
  \sum_{i=1}^{N} P_{i,j} \Dpartial{\mJ}{\sigma_i} \Delta_i.
  \label{eq:grad_zeta}
\end{equation}
In general, a gradient-based framework to perform optimization on multiple,
fine and coarse, meshes will benefit from the following.
\begin{itemize}
  \item Forward simulations on fine N-element meshes allow
    constructing highly accurate adjoint-based gradients
    $\bnabla_{\sigma} \mJ$.
  \item Reasonably small number of controls $N_{\zeta} \ll N$ defined at
    coarse scales tends to lessen the number of local minima.
  \item Upscaling gradients at coarse scales by \eqref{eq:grad_zeta}
    allows dynamical control space relaxation without interrupting iterative
    optimization to follow changes in upscaling map \eqref{eq:upscale_map}.
\end{itemize}
Figure~\ref{fig:mpe_eit} illustrates the general concept of such dynamical
control space relaxation and obtaining upscaled gradients for two new controls
at a coarse scale. As shown in Figures~\ref{fig:mpe_eit}(a,b), gradient
components for all fine mesh controls may have the same order of magnitude.
Thus, these controls, shown in red and blue, could be grouped following
the idea which is different from the analysis of the gradient structure
used in GBME. We discuss this in detail as a new grouping (partitioning)
approach in Section~\ref{sec:OPTframe_coarse}.

\section{Multiscale Optimization Framework}
\label{sec:OPTframe}

\subsection{Switching Between Scales}
\label{sec:OPTframe_general}

\begin{figure}[!htb]
  \begin{center}
  \includegraphics[width=1.0\textwidth]{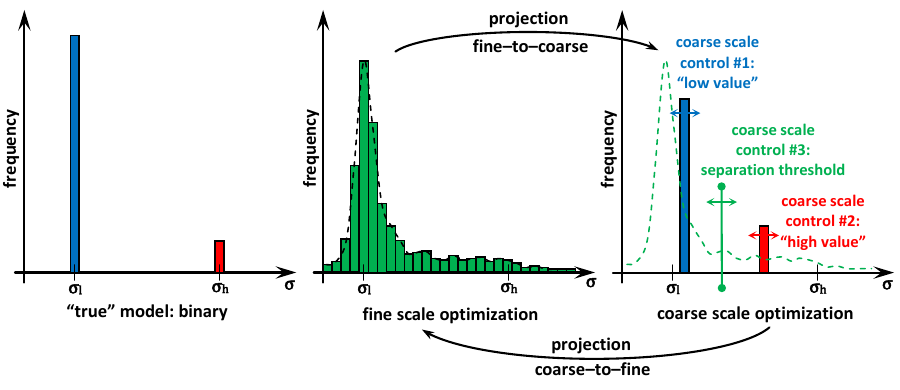}
  \end{center}
  \caption{Schematic illustrating the general concept of the multiscale
    optimization framework. (left)~A typical histogram representing
    binary distribution of true electrical conductivity $\sigma(x)$ used
    in EIT. In all three plots $\sigma_l$ and $\sigma_h$ values represent
    two modes associated respectively with healthy and cancer-affected
    regions within domain $\Omega$.
    (middle)~An example of the Gaussian-type histogram typical for solution
    $\sigma^k(x)$ obtained after $k$ iterations at a fine scale.
    (right)~A binary histogram for solution $\sigma^k(x)$ obtained at $k$-th
    iteration at a coarse scale. Positions of blue and red bars are associated
    with current values of $\sigma^k_{low}$ and $\sigma^k_{high}$ controls, while
    their heights are computed based on the fine scale representation
    $\sigma(\xi^k)$ cut off by the current value of the coarse scale
    threshold control $\sigma^k_{th}$.
    See Sections~\ref{sec:OPTframe_fine}--\ref{sec:OPTframe_coarse} for
    details. Coarse--to--fine and fine--to--coarse projections are defined
    respectively by \eqref{eq:proj_CtoF}--\eqref{eq:proj_CtoF_rlx} and
    \eqref{eq:sigma_coarse}--\eqref{eq:minJ_zeta_ini_2}.}
  \label{fig:mpe_cartoon}
\end{figure}

The proposed approach for optimization utilizing multilevel control space
reduction over multiple scales, both fine and coarse, is motivated by
a range of possible applications in biomedical sciences. These applications
are based on physical models represented by binary distributions of some
physical properties, e.g.~electrical conductivity $\sigma(x)$ in EIT.
Figure~\ref{fig:mpe_cartoon}(left) shows an example of the histogram
typical for such distributions.

A common approach to solve optimization problem \eqref{eq:minJ_sigma}
is to solve problem \eqref{eq:minJ_xi} instead by applying PCA-based space
reduction by mapping $N$-element (fine-mesh) $\sigma$-space and
reduced-dimensional $\xi$-space. An optimal solution $\hat \sigma(x)$
obtained after applying map \eqref{eq:map_pca_1} to $\hat \xi$ is of
a Gaussian type. In case one of two modes is relatively small, a histogram
for the solution image is hardly recognized as being bimodal, e.g.~as shown
in Figure~\ref{fig:mpe_cartoon}(middle), and possible conversions to binary
images may be very inaccurate.

To provide a remedy and obtain the optimal solution $\hat \sigma(x)$
of a required binary type the proposed approach employs multiscale
optimization at both fine and coarse scales each with their own sets
of controls by using them interchangeably. While various schemes are
available for switching between scales, here we consider a simple one:
scales are changed every $n_s$ optimization iterations. We also define
the coarse scale indicator function
\begin{equation}
  \chi_c (k) = \left\{
  \begin{aligned}
    0, \quad (2k_s - 2) n_s &< k \leq (2k_s - 1) n_s,
    && \quad {\rm(fine \ scale)}\\
    1, \quad (2k_s - 1) n_s &< k \leq 2k_s n_s,
    && \quad {\rm(coarse \ scale)}
  \end{aligned}
  \right.
  \label{eq:scale_ind}
\end{equation}
where $k_s = 1, 2, \ldots$ and $k = 0, 1, 2, \ldots$ denote respectively
the counts for switching cycles and optimization iterations.
We choose to terminate the optimization run once the following
criterion is satisfied
\begin{equation}
    \left| \dfrac{\mJ(\sigma^k) - \mJ(\sigma^{k-1})}{\mJ(\sigma^k)}
    \right| < (1-\chi_c) \epsilon_f + \chi_c \epsilon_c,
    \quad k \neq k_s n_s + 1
  \label{eq:termination}
\end{equation}
subject to chosen tolerances $\epsilon_f, \epsilon_c \in \RR_+$.

Our multiscale optimization framework is shown schematically in
Figures~\ref{fig:mpe_cartoon}(middle,right). Here, we would like to
emphasize that all solutions obtained at both fine and coarse scales are
represented on (projected onto) the same $N$-element fine mesh. The word
``multiscale'', in fact, refers to updates provided to discretized control
$\sigma(x)$. At a fine scale all elements $\sigma_i, \, i = 1, \ldots, N$,
are updated by applying PCA-based transformation, while at
a coarse scale these elements are sorted into two groups and
then updated within each group by following the same rule. We discuss
optimization phases at both scales as well as switching procedures
in the following two sections.

\subsection{Fine Scale Phase}
\label{sec:OPTframe_fine}

We denote the solution for control $\sigma^k = \sigma^k(x)$ obtained
at a fine scale at $k$-th iteration as $\sigma(\xi^k)$. During the
fine scale optimization phase, $\chi_c(k) = 0$, control $\sigma^k$
is updated by solving optimization problem \eqref{eq:minJ_xi}
in the reduced-dimensional $\xi$-space and by using map \eqref{eq:map_pca_1}
as described in Section~\ref{sec:fine_mesh}, i.e.~$\sigma^k = \sigma(\xi^k)$.
Alternatively, during the coarse scale optimization phase ($\chi_c(k) = 1$)
$\sigma(\xi^k)$, updated last time at the end of the fine scale phase,
is used in coarse scale control grouping discussed in
Section~\ref{sec:OPTframe_coarse}.

To start the first switching cycle, $k_s = 1$, and optimization itself,
$k = 0$, the initial guess $\sigma^0$ may be taken, for instance,
corresponding to any approximate theoretical prediction $\sigma_0(x)$.
It is obvious that every time when coarse scale switches to the fine one,
$k = 2k_s n_s, \, k_s \geq 1$, fine scale controls $\xi^k$ should be
updated to ensure it receives as much as possible information related
to recent changes in $\sigma^k$ during the coarse scale phase. On the
other hand, this update should not worsen the results $\sigma(\xi^k)$
previously obtained at the fine scale. Here, we propose the following
scheme to project solution $\sigma^k$ obtained at the end of the coarse
scale phase onto $\xi$-space by using a convex combination of
$\sigma(\xi^k)$ and $\sigma^k$
\begin{equation}
  \forall k = 2k_s n_s, \, k_s \geq 1 : \ \
  \bar \sigma(\xi^k) = \alpha_{c \rightarrow f} \, \sigma(\xi^k)
  + (1 - \alpha_{c \rightarrow f}) \, \sigma^k, \quad
  \alpha_{c \rightarrow f} \in [0,1].
  \label{eq:proj_CtoF}
\end{equation}
Control $\xi^k$ then could be re-initialized from $\bar \sigma(\xi^k)$
by using map \eqref{eq:map_pca_2}. As $\bar \sigma(\xi^k)$ and
$\sigma^k$ have different distributions, namely of Gaussian and binary
types, the coarse--to--fine projection scheme \eqref{eq:proj_CtoF} may
benefit from projecting $\sigma^k$ first to its PCA equivalent
\begin{equation}
  \sigma^k_{PCA} = \Phi \hat \Phi^{-1} (\sigma^k - \bar \sigma)
  + \bar \sigma,
  \label{eq:proj_CtoF_pcaProj}
\end{equation}
see \cite{VolkovBukshtynov18,Bukshtynov15} for details.
$\sigma^k_{PCA}$ then could be used in \eqref{eq:proj_CtoF}
in place of $\sigma^k$. Also, an optimal choice of relaxation
parameter $\alpha_{c \rightarrow f}$ could be made by solving the
following optimization problem in 1D
\begin{equation}
  \alpha_{c \rightarrow f} =
  \hat \alpha =
  \underset{
  \begin{array}{cc}
     0 \leq \alpha \leq 1\\
     \mJ(\bar \sigma(\xi^k)) \leq \mJ(\sigma(\xi^k))
  \end{array}
  }{\argmin \ \alpha}
  \label{eq:proj_CtoF_rlx}
\end{equation}
which appears to be highly nonlinear due to the inequality constraint
to control the quality of fine scale solutions $\sigma(\xi^k)$
in transition between subsequent switching cycles.

\subsection{Coarse Scale Phase}
\label{sec:OPTframe_coarse}

To run optimization at a coarse scale we define a new 3-component
control vector $\zeta = (\zeta_j)_{j=1}^3$ in which the first two
entries are the low and high values of (binary) electrical conductivity
$\sigma (x)$ associated with healthy and cancer-affected regions
in domain $\Omega$, i.e.
\begin{equation}
  \zeta_1 = \sigma_{low}, \quad \zeta_2 = \sigma_{high}.
  \label{eq:ctrls_coarse_1_2}
\end{equation}
They are shown schematically as respectively blue and red bars
in Figure~\ref{fig:mpe_cartoon}(right). The third component,
$\zeta_3 = \sigma_{th}$, takes responsibility for the shape of those
regions (healthy and cancer-affected) and is set as a separation
threshold to define a boundary between low and high conductivity
regions as shown in green in Figure~\ref{fig:mpe_cartoon}(right).
Such a simple structure of control $\zeta$ allows us to create
a simplified representation of the coarse scale solution $\zeta^k$
for control $\sigma^k$ at $k$-th iteration based on the current fine
scale representation $\sigma(\xi^k) = (\sigma_i(\xi^k))_{i=1}^N$
\begin{equation}
  \sigma^k_i = \left\{
  \begin{aligned}
    \sigma^k_{low},  & \quad \sigma_i(\xi^k) < \sigma^k_{th},\\
    \sigma^k_{high}, & \quad \sigma_i(\xi^k) \geq \sigma^k_{th},
  \end{aligned}
  \right. \qquad i = 1, \ldots N,
  \label{eq:sigma_coarse}
\end{equation}
where
\begin{equation}
  0 < \sigma^k_{low} < \sigma^k_{high}, \qquad
  \underset{i}{\min} \ \sigma_i(\xi^k) < \sigma^k_{th} <
  \underset{i}{\max} \ \sigma_i(\xi^k), \ i = 1, \ldots N.
  \label{eq:sigma_coarse_bounds}
\end{equation}
Simply, \eqref{eq:sigma_coarse} provides a rule for creating
fine--to--coarse partition $\mM$ in \eqref{eq:upscale_map} when
$N_{\zeta} = 2$ based on the current state of control $\zeta$.
During the coarse scale optimization phase, $\chi_c(k) = 1$,
control $\sigma^k$ is updated by solving a 3D optimization problem
in the $\zeta$-space
\begin{equation}
  \hat \zeta =
  \underset{\zeta}{\argmin} \ \mJ(\zeta)
  \label{eq:minJ_zeta}
\end{equation}
subject to constraints (bounds) provided in
\eqref{eq:sigma_coarse_bounds}, and then $\sigma^k = \sigma(\zeta^k)$.
When solving problem \eqref{eq:minJ_zeta} during the first switching cycle,
$k = n_s$, $\zeta^k$ could be initially approximated by some constants,
for example
\begin{equation}
  \begin{aligned}
    \sigma^k_{th} &= \frac{1}{2}
    \left[ \underset{i}{\max} \ \sigma_i(\xi^k) +
    \underset{i}{\min} \ \sigma_i(\xi^k) \right],\\
    \sigma^k_{low} &= \underset{i}{\mean}
   \left\{ \sigma_i(\xi^k) : \ \sigma_i(\xi^k) < \sigma^k_{th}
   \right\},\\
    \sigma^k_{high} &= \underset{i}{\mean}
    \left\{ \sigma_i(\xi^k) : \ \sigma_i(\xi^k) \geq \sigma^k_{th}
    \right\}, \quad i = 1, \ldots N.
  \end{aligned}
  \label{eq:minJ_zeta_ini_1}
\end{equation}
Switching from fine scale to coarse one when
$k = (2k_s - 1) n_s, \ k_s > 1$, could be even more straightforward
by utilizing the components of control $\zeta$ obtained at the
previous coarse scale phase, i.e.
\begin{equation}
  \zeta^k = \zeta^{k-2n_s}.
  \label{eq:minJ_zeta_ini_2}
\end{equation}
In order to solve \eqref{eq:minJ_zeta} by any approach which
requires computing a gradient, its first two components
\begin{equation*}
  \Dpartial{\mJ(\zeta)}{\zeta_1} = \Dpartial{\mJ}{\sigma_{low}}, \quad
  \Dpartial{\mJ(\zeta)}{\zeta_2} = \Dpartial{\mJ}{\sigma_{high}}
\end{equation*}
could be easily obtained by using gradient summation formula
\eqref{eq:grad_zeta} after completing partitioning map $\mM$
\eqref{eq:upscale_map}--\eqref{eq:part_ind} by employing
\eqref{eq:sigma_coarse}. On the other hand, the third component
may be approximated by a finite difference scheme,
e.g.~of the first order,
\begin{equation}
  \Dpartial{\mJ(\zeta)}{\zeta_3} = \Dpartial{\mJ}{\sigma_{th}} =
  \dfrac{\mJ \left(\sigma^k(\zeta_1,\zeta_2,\zeta_3 +
  \delta_{\zeta}) \right) - \mJ \left( \sigma^k(\zeta_1,\zeta_2,\zeta_3)
  \right)} {\delta_{\zeta}} + \mO(\delta_{\zeta}).
  \label{eq:grad_zeta_3}
\end{equation}
Parameter $\delta_{\zeta}$ in \eqref{eq:grad_zeta_3} is to be set
experimentally pursuing trade-off between being reasonably small to
ensure accuracy and large enough to protect numerator from being zero.
In fact, formulas \eqref{eq:sigma_coarse}--\eqref{eq:minJ_zeta_ini_2}
provide a complete description of the fine--to--coarse projection for
control $\sigma(x)$ used in our approach. A summary of the complete
computational scheme to perform our new PCA-based multilevel
optimization over multiple scales is provided in
Algorithm~\ref{alg:main_opt}.

\begin{algorithm}[ht!]
\begin{algorithmic}
  \STATE $k \leftarrow 0$
  \STATE $\chi_c \leftarrow 0$
  \STATE $\sigma^0 \leftarrow $ initial guess $\sigma_0(x)$
  \STATE compute $\xi^0$ using $\sigma^0$ by \eqref{eq:map_pca_2}
  \REPEAT
  \STATE compute $u^k$ using $\sigma^k$ by solving forward problem
    \eqref{eq:forward}
  \STATE compute $\psi^k$ using $u^k$ and $\sigma^k$ by solving adjoint
    problem \eqref{eq:adjoint}
  \STATE compute $\bnabla_{\sigma} \mJ(\sigma^k)$ using $u^k$ and $\psi^k$
    by \eqref{eq:grad_sigma}
  \IF{$\chi_c = 1$}
    \STATE compute $\sigma(\xi^k)$ using $\xi_k$ by \eqref{eq:map_pca_1}
    \STATE compute $\bnabla_{\zeta} \mJ(\zeta^k)$ using $\zeta^k,
      \sigma(\xi^k),$ and $\bnabla_{\sigma} \mJ(\sigma^k)$ by
      \eqref{eq:grad_zeta},
      \eqref{eq:upscale_map}--\eqref{eq:part_ind},
      \eqref{eq:sigma_coarse}, and \eqref{eq:grad_zeta_3}
  \ELSE
    \STATE compute $\bnabla_{\xi} \mJ(\xi^k)$ using $\bnabla_{\sigma}
      \mJ(\sigma^k)$ by \eqref{eq:grad_xi}
  \ENDIF
  \STATE update $\zeta^{k+1}$ and $\xi^{k+1}$ by using, depending
    on $\chi_c(k)$, descent directions $D_{\zeta}\left( \bnabla_{\zeta}
    \mJ\right)$ or $D_{\xi}\left( \bnabla_{\xi} \mJ \right)$
    obtained respectively from $\bnabla_{\zeta} \mJ(\zeta^k)$ or
    $\bnabla_{\xi} \mJ(\xi^k)$
  \begin{subequations}
    \begin{alignat}{1}
      \zeta^{k+1} &= \zeta^k - \chi_c(k) \tau^k D_{\zeta}
        \left( \bnabla_{\zeta} \mJ(\zeta^k) \right),
        \label{eq:it_zeta}\\
      \xi^{k+1} &= \xi^k - (1-\chi_c(k)) \tau^k D_{\xi}
        \left( \bnabla_{\xi} \mJ(\xi^k) \right) \label{eq:it_xi}
    \end{alignat}
    \label{eq:it_zeta_xi}
  \end{subequations}
  \IF{$\chi_c = 1$}
    \STATE compute $\sigma^{k+1}$ using $\zeta^{k+1}$ and $\sigma(\xi^{k+1})$
      by \eqref{eq:sigma_coarse}
  \ELSE
    \STATE compute $\sigma^{k+1}$ using $\xi^{k+1}$ by
      \eqref{eq:map_pca_1}
  \ENDIF
  \STATE $k \leftarrow k + 1$
  \STATE update $\chi_c$ using $k$ by \eqref{eq:scale_ind}
  \IF{$\chi_c(k) \neq \chi_c(k-1)$}
    \IF{$\chi_c = 1$}
      \STATE update $\sigma^k$ using $\zeta^k$ and $\sigma(\xi^k)$ by
        \eqref{eq:sigma_coarse}
    \ELSE
      \STATE update $\xi^k$ using $\sigma^k$ and $\sigma(\xi^k)$ by
        \eqref{eq:proj_CtoF}--\eqref{eq:proj_CtoF_rlx}
      \STATE update $\sigma^k$ using $\xi^k$ by \eqref{eq:map_pca_1}
    \ENDIF
  \ENDIF
  \UNTIL termination criterion \eqref{eq:termination} is satisfied
    to given tolerances $\epsilon_f$ and $\epsilon_c$
\end{algorithmic}
\caption{Computational workflow for PCA-based multiscale
  optimization}
\label{alg:main_opt}
\end{algorithm}

\section{Main Results}
\label{sec:results}

\subsection{Computational Model in 2D}
\label{sec:comp_model}

\begin{figure}[!htb]
  \begin{center}
  \mbox{
  \subfigure[]{\includegraphics[width=0.33\textwidth]{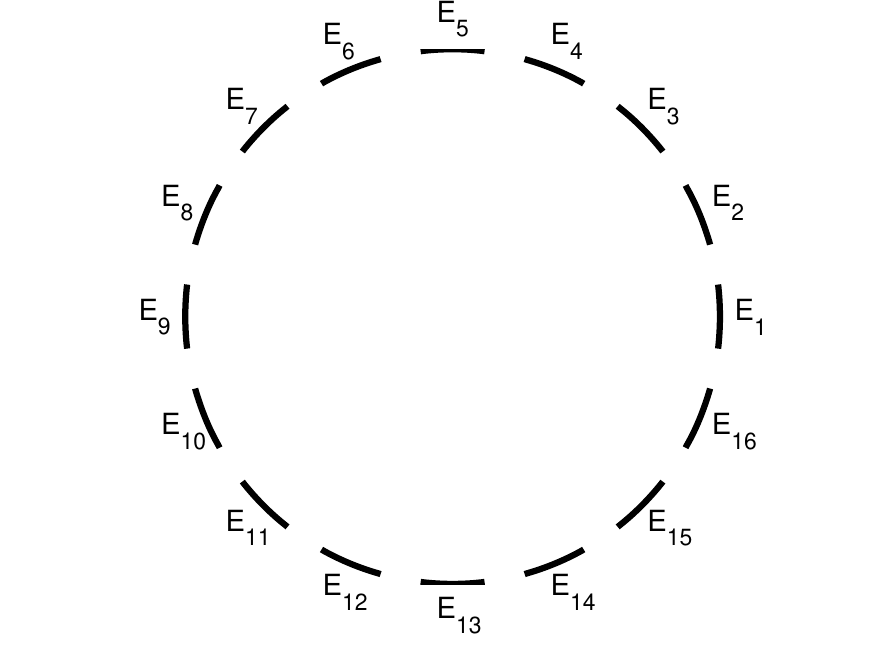}}
  \subfigure[]{\includegraphics[width=0.33\textwidth]{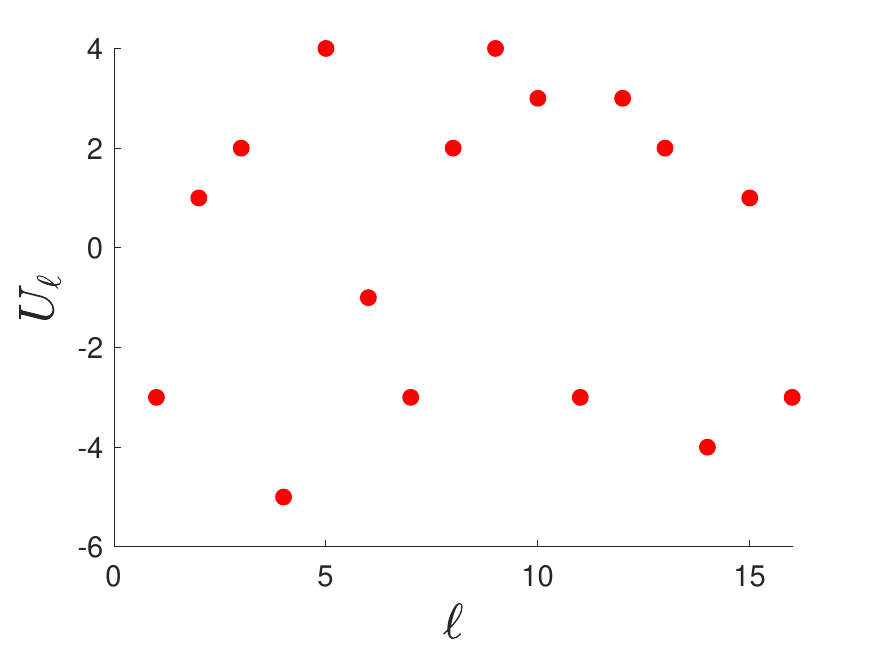}}
  \subfigure[]{\includegraphics[width=0.33\textwidth]{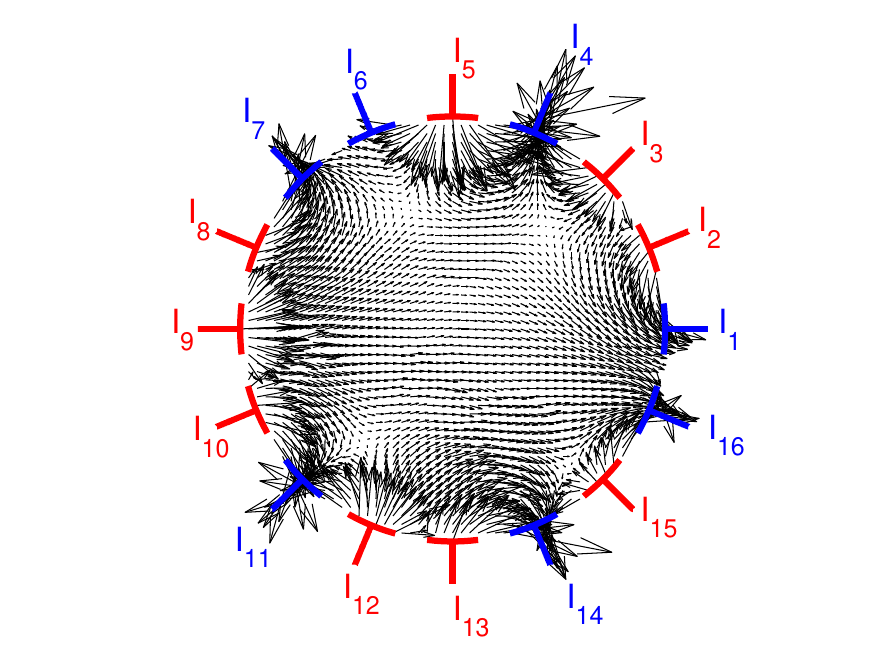}}}
  \end{center}
  \caption{(a)~Equispaced geometry of electrodes $E_{\ell}, \ \ell = 1, 2, \ldots, 16$,
    placed over boundary~$\partial \Omega$. (b)~Electrical potentials $U_{\ell}$.
    (c)~Electrical currents $I_{\ell}$ (positive in red, negative in blue)
    measured at electrodes $E_{\ell}$. Black arrows show the distribution
    of flux $\sigma(x) \bnabla u(x)$ of electrical potential $u(x)$ in the
    interior of domain $\Omega$.}
  \label{fig:model}
\end{figure}

Our optimization framework integrates computational facilities for
solving forward PDE problem \eqref{eq:forward}, adjoint PDE problem
\eqref{eq:adjoint}, and evaluation of the gradients according to
\eqref{eq:grad_sigma}, \eqref{eq:grad_xi}, and \eqref{eq:grad_zeta}.
These facilities are incorporated mainly by using {\tt FreeFem++},
see \cite{FreeFem2012} for details, an open--source, high--level
integrated development environment for obtaining numerical solutions
of PDEs based on the Finite Element Method (FEM). For solving
numerically forward PDE problem \eqref{eq:forward}, spatial
discretization is carried out by implementing FEM triangular finite
elements: P2 piecewise quadratic (continuous) representation for
electrical potential $u(x)$ and P0 piecewise constant representation
for conductivity field $\sigma(x)$. Systems of algebraic equations
obtained after such discretization are solved with {\tt UMFPACK},
a solver for nonsymmetric sparse linear systems \cite{UMFPACK}.
The same technique is used for numerical solutions of adjoint
problem \eqref{eq:adjoint}. All computations are performed using
2D domain
\begin{equation}
  \Omega = \left\{ x \in \RR^2 : \ x_1^2 + x_2^2 < r^2_{\Omega}
  \right\}
  \label{eq:domain2D}
\end{equation}
which is a disc of radius $r_{\Omega} = 0.1$ with $m = 16$
equidistant electrodes $E_{\ell}$ with half-width $w = 0.12$ rad
covering approximately 61\% of boundary $\partial \Omega$ as shown
in Figure~\ref{fig:model}(a). Electrical potentials $U_{\ell}$, see
Figure~\ref{fig:model}(b), are applied to electrodes $E_{\ell}$ as
seen in \eqref{eq:permutations} following the ``rotation scheme''
discussed in Section~\ref{sec:model}. We also consider adding up to
three additional permutations within the set of potentials $U$,
and, by choosing $K \in \left\{ 1, 2, 3, 4 \right\}$, we increase
the total number of measurements from $m^2 = 256$ ($K=1$) to $K m^2 = 1024$
($K = K_{\max} = 4$). The potentials are chosen to be consistent with
the ground potential condition \eqref{eq:curr_volt_conds}. Determining
the Robin part of the boundary conditions in \eqref{eq:forward_3} we
equally set the electrode contact impedance $Z_{\ell} = 0.1$.
Figure~\ref{fig:model}(c) also shows an example of the distribution
of flux $\sigma(x) \bnabla u(x)$ of electrical potential $u(x)$ in
the interior of domain $\Omega$ during EIT.

Physical domain $\Omega$ is discretized using mesh created by specifying
$176$ vertices over boundary $\partial \Omega$ and totaling $N = 7726$
triangular FEM elements inside $\Omega$. This fine mesh is then used to
construct gradients $\bnabla_{\sigma} \mJ$, $\bnabla_{\xi} \mJ$, and
$\bnabla_{\zeta} \mJ$ to perform the optimization procedure as described
in Algorithm~\ref{alg:main_opt}. To solve problems \eqref{eq:minJ_zeta}
and \eqref{eq:minJ_xi} iteratively as seen in \eqref{eq:it_zeta_xi}, our
framework is utilizing respectively the Steepest Descent~(SD) and
Conjugate Gradient~(CG) approaches \cite{Nocedal2006} to obtain descent
directions $D_{\zeta}$ and $D_{\xi}$. Stepsize parameters $\tau^k$ in
\eqref{eq:it_zeta_xi} are obtained by applying line minimization search
\cite{NumericalRecipes2007}.

The actual (true) electrical conductivity $\sigma_{true}(x)$ we seek to
reconstruct will be given analytically for each model by
\begin{equation}
  \sigma_{true}(x) = \left\{
  \begin{aligned}
    \sigma_c, & \quad x \in \Omega_c,\\
    \sigma_h, & \quad x \in \Omega_h,
  \end{aligned}
  \right.
  \qquad \Omega = \Omega_c \cup \Omega_h,
  \qquad \Omega_c \cap \Omega_h = \emptyset
  \label{eq:sigma_true}
\end{equation}
and setting $\sigma_c = 0.4$ for cancer-affected region $\Omega_c$
(up to 4 spots of different size depending on the model's complexity)
and $\sigma_h = 0.2$ to healthy tissues part $\Omega_h$. In terms of
the initial guess for control $\sigma(x)$ we take a constant
approximation to $\sigma_{true}$ given by
$\sigma_0 = \frac{1}{2} \left(\sigma_h + \sigma_c \right) = 0.3$.
In order to avoid early termination at a coarse scale, termination
tolerances in \eqref{eq:termination} are set as $\epsilon_c = 0$ and
$\epsilon_f = 10^{-10}$.

To simplify the enforcement of bounds established for coarse scale
control $\sigma^k_{th}$ in \eqref{eq:sigma_coarse_bounds}, in all computations
we used fine--to--coarse partition \eqref{eq:sigma_coarse} redefined as
\begin{equation}
  \sigma^k_i = \left\{
  \begin{aligned}
    \sigma^k_{low},  & \quad \sigma_i(\xi^k) < (1-\sigma^k_{th}) \,
    \underset{i}{\min} \ \sigma_i(\xi^k) + \sigma^k_{th} \,
    \underset{i}{\max} \ \sigma_i(\xi^k),\\
    \sigma^k_{high}, & \quad {\rm otherwise},
  \end{aligned}
  \right. \qquad i = 1, \ldots N,
  \label{eq:sigma_coarse_new}
\end{equation}
while ensuring $0 < \sigma^k_{th} < 1$.

For all computations mentioned in the rest of Section~\ref{sec:results}
we use a PCA-based map \eqref{eq:map_pca} between $N$-dimensional
discretized control $\sigma(x)$ and reduced-dimensional $\xi$-space as
described in Section~\ref{sec:fine_mesh}. A set of $N_r = 1000$
realizations $(\sigma^*_n)_{n=1}^{1000}$ is created using a generator
of uniformly distributed random numbers. Each realization $\sigma^*_n$
``contains'' from one to seven ``cancer-affected'' areas with
$\sigma_c = 0.4$. Each area is located randomly within domain $\Omega$
and represented by a circle of randomly chosen radius
$0 < r \leq 0.3 r_{\Omega}$. To perform TSVD we choose the number of
principal components $N_{\xi}$ by retaining 662, 900, and 965 basis
vectors in the PCA description. These values correspond to the
preservation of respectively $r_{\xi} = 99\%$, $99.9\%$, and $99.99\%$
of the ``energy'' in the full set of basis vectors, see
\cite{AbdullaBukshtynovSeif,Bukshtynov15} for details.

\subsection{Parameter Calibration}
\label{sec:model_1}

\begin{figure}[!htb]
  \begin{center}
  \mbox{
  \subfigure[]{\includegraphics[width=0.33\textwidth]{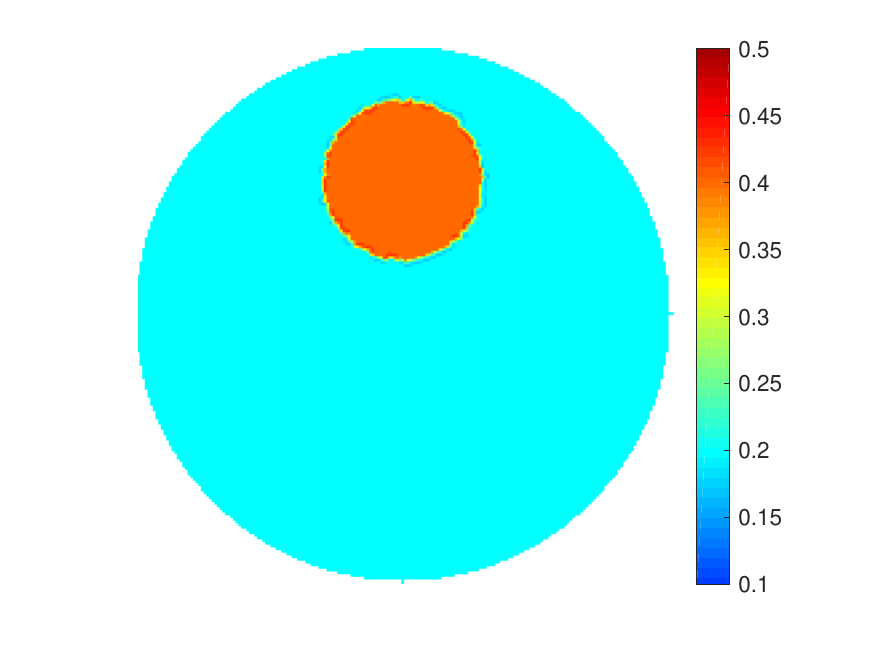}}
  \subfigure[]{\includegraphics[width=0.33\textwidth]{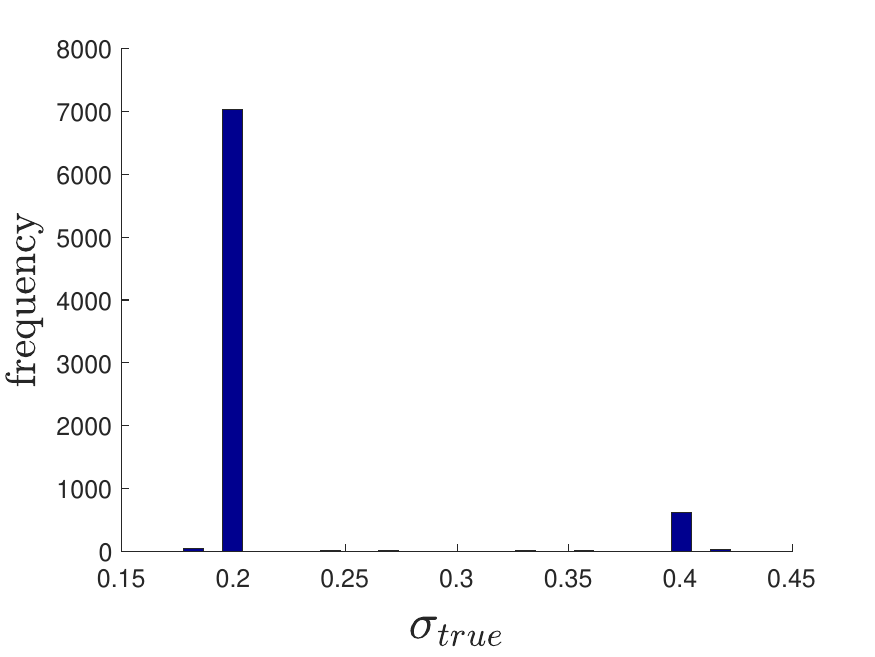}}
  \subfigure[]{\includegraphics[width=0.33\textwidth]{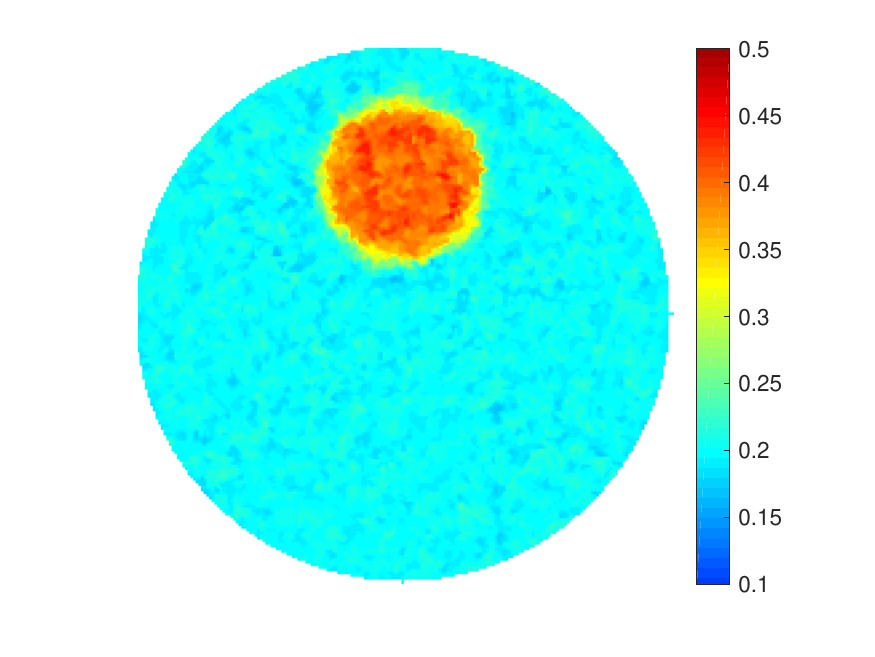}}}
  \end{center}
  \caption{Model \#1. (a)~True electrical conductivity $\sigma_{true}(x)$.
    (b)~A binary-type histogram for model \#1. (c)~Electrical conductivity
    $\sigma_{PCA}(x)$ for model \#1 obtained by projecting $\sigma_{true}(x)$
    to its PCA equivalent with $r_{\xi} = 99\%$ by using \eqref{eq:proj_CtoF_pcaProj}.}
  \label{fig:model_9}
\end{figure}

We created our (benchmark) model \#1 to check the performance of the
proposed optimization framework and discuss the procedure for calibrating
its certain parameters. This model in fact mimics a simple situation
when a biological tissue contains one circular-shaped area suspicious
to be affected by cancer as seen in Figures~\ref{fig:model_9}(a,b).

Here, we would like to address a well-known issue on the presence of noise in
measurements due to improper electrode--tissue contacts, wire interference,
possible electrode misplacement, etc. Its negative impact has been already
investigated by many researchers both theoretically and within practical
applications. Although all results in this paper are obtained without explicit
noise added to measurements, our synthetic data features effects similar
to those when noise is present. First, ``measured'' electrical currents
$I^*_{\ell}$ are recorded by running \eqref{eq:forward} and \eqref{eq:el_current}
with $\sigma(x) = \sigma_{true}$ represented by P2 FEM elements. Then,
as mentioned in Section~\ref{sec:comp_model}, the
actual reconstruction $\hat \sigma(x)$ is obtained in P0 finite element
space. Taking also into account that the ``measurement'' data is not projected
into its PCA equivalent, we compared cost functional $\mJ(\sigma)$ evaluated
at $\sigma_{true}$ with added noise and at $\sigma_{PCA}$, as shown in
Figure~\ref{fig:model_9}(c), obtained after applying PCA projection.
The equivalent noise is estimated to be at level $0.2\%-0.3\%$.

In order to evaluate the performance, we define a set
$\mS = \left\{ K, \ r_{\xi}, \ j_{\max}, \ n_s \right\}$ of 4 major
parameters to be calibrated:
\begin{itemize}
  \item $K \in \{1, 2, 3, 4\}$ to define the total number of measurements,
    $K m^2$, respectively as 256, 512, 768, 1024,
  \item $r_{\xi} \in \{99\%, 99.9\%, 99.99\%\}$ to set the number of
    principal components, i.e.~number of controls $N_{\xi}$ at a fine scale,
    respectively to 662, 900, 965,
  \item $j_{\max} \in \{ 1, Km \}$ to consider two cases for optimization at
    the coarse scale: $\beta^j_{\ell} = 1, \, j = 1, \ldots, Km$, vs.~$\beta^j_{\ell} = 0$
    except $j=1$ to consider respectively full data vs.~$m = 16$ measurements
    to avoid over-parameterization, and
  \item $n_s \in \{5, 10\}$ frequency of switching between fine and coarse
    scales.
\end{itemize}
\begin{figure}[!htb]
  \begin{center}
  \mbox{
  \subfigure[]{\includegraphics[width=0.5\textwidth]{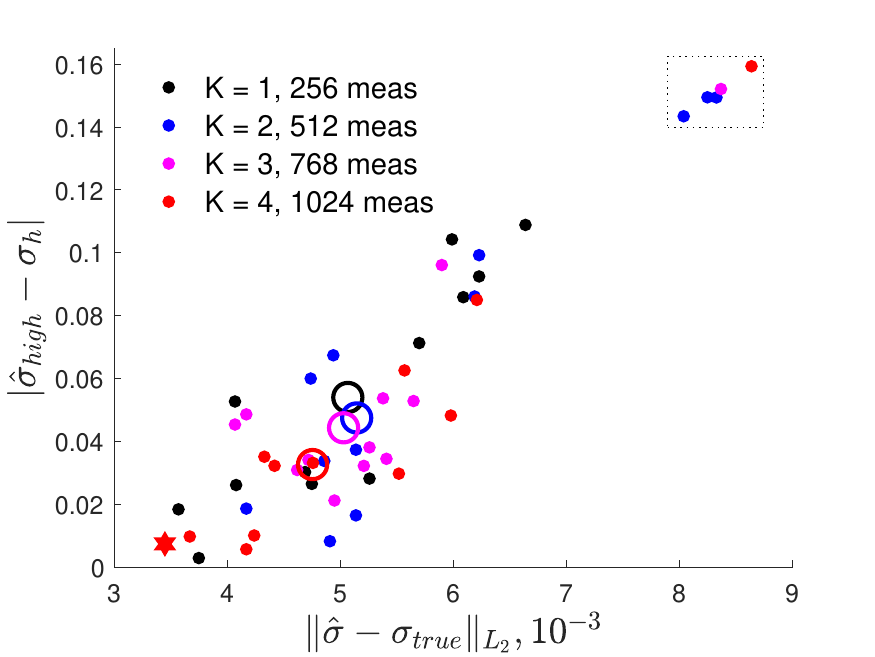}}
  \subfigure[]{\includegraphics[width=0.5\textwidth]{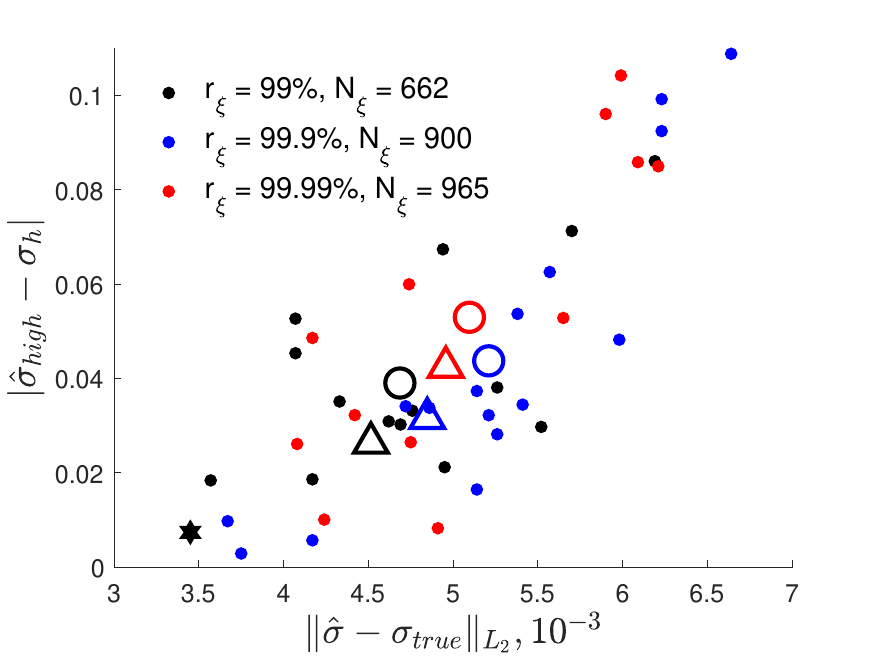}}}
  \mbox{
  \subfigure[]{\includegraphics[width=0.5\textwidth]{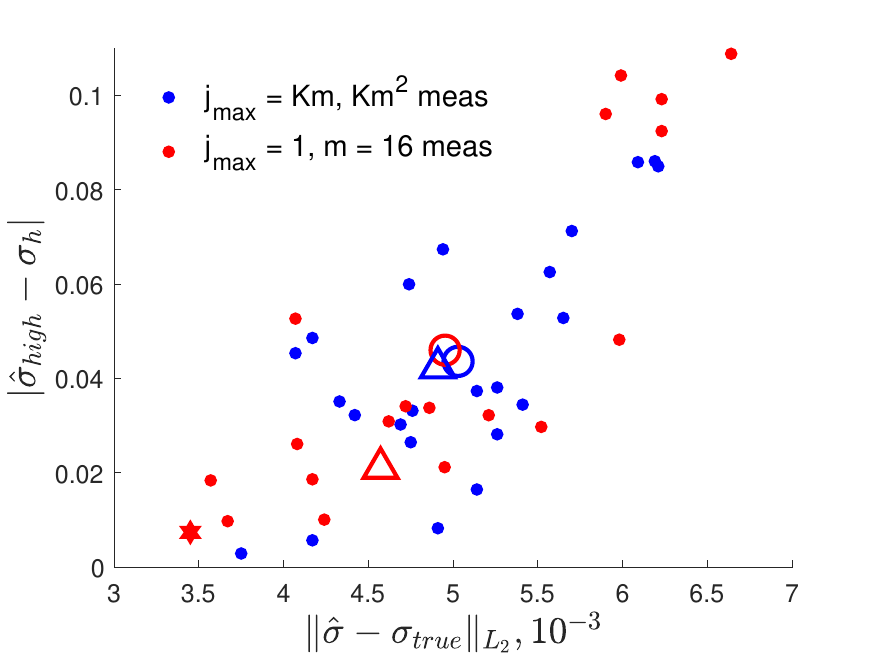}}
  \subfigure[]{\includegraphics[width=0.5\textwidth]{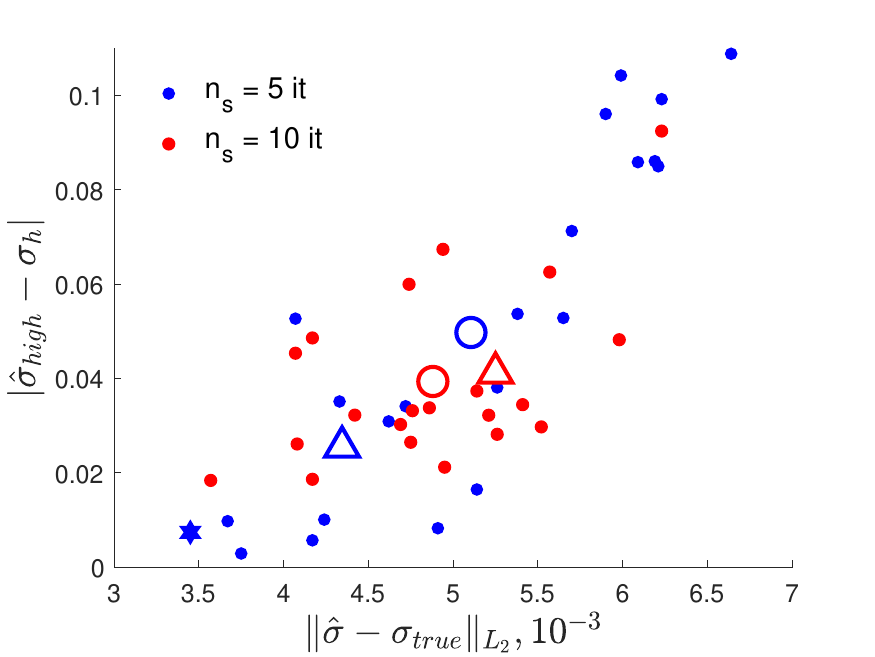}}}
  \end{center}
  \caption{Model \#1 calibration results. Different colors are used for different
    values of parameters (a)~$K = 1, 2, 3, 4$, (b)~$r_{\xi} = 99\%, 99.9\%,
    99.99\%$, (c)~$j_{\max} = 1, Km$, and (d)~$n_s = 5, 10$. Open circles in
    (a) provide the results averaged over four cases $K = 1, 2, 3, 4$.
    Open circles and triangles in (b,c,d) represent averages obtained
    respectively for $K = 1$ and $K = 4$. Five outliers shown inside the
    dotted box in (a) are removed from plots in (b,c,d).
    The result obtained with parameter schedule $\mS^*$ in
    \eqref{eq:opt_schedule} is shown by stars in all four plots.}
  \label{fig:model_9_calibr}
\end{figure}

Figure~\ref{fig:model_9_calibr} shows the results obtained after performing
the calibration procedure, i.e.~after running optimization for model \#1 with
48 different parameter schedules $\mS$. We have chosen two critical factors to
evaluate the performance in each case: the $L_2$-norm difference between
optimal solution $\hat \sigma(x)$ and true conductivity field $\sigma_{true}(x)$
($x$-axis), and the absolute error in recovered $\hat \sigma_{high}$ by
comparing it with the known value $\sigma_h = 0.4$ ($y$-axis). We note that for
all experiments $\hat \sigma_{low}$ is recovered very accurately.
Five outliers inside the dotted box in Figure~\ref{fig:model_9_calibr}(a)
are excluded from the entire calibration statistics. Then open circles in
Figure~\ref{fig:model_9_calibr}(a) provide the results averaged over four
cases, $K = 1, 2, 3, 4$, with $K = 4$ (red circle) appeared as the best one.
Figures~\ref{fig:model_9_calibr}(b,c,d) show the same $43$ outcomes (less
5 outliers) then colored according to values of the rest parameters, namely
$r_{\xi}, j_{\max}$, and $n_s$. Open circles and triangles there represent
averages obtained respectively for $K = 1$ and $K = 4$. We conclude that
simple models, like our model~\#1, will be best reconstructed with
99\% PCA, limited data at a coarse scale, and by switching between
scales every 5 iterations, i.e.~our calibration returns the following
schedule
\begin{equation}
  \mS^* = \left\{ K = 4, \ r_{\xi} = 99\%, \ j_{\max} = 1, \ n_s = 5 \right\}.
  \label{eq:opt_schedule}
\end{equation}
\begin{figure}[!htb]
  \begin{center}
  \mbox{
  \subfigure[]{\includegraphics[width=0.33\textwidth]{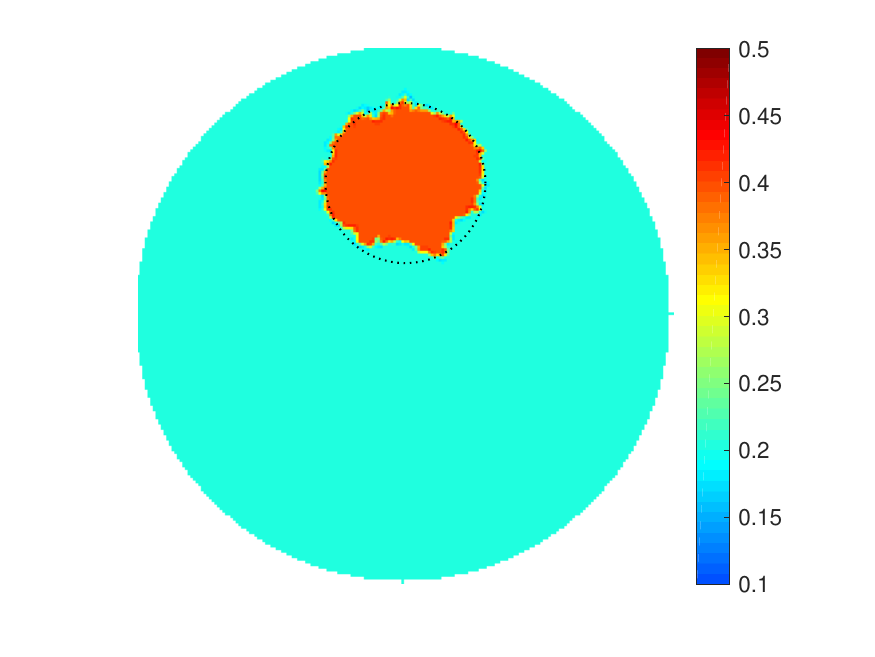}}
  \subfigure[]{\includegraphics[width=0.33\textwidth]{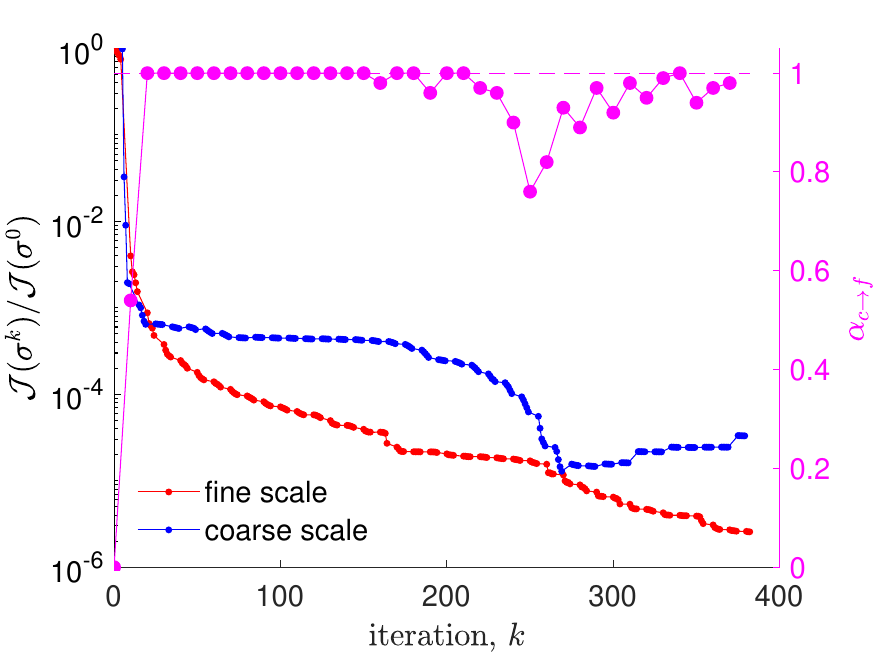}}
  \subfigure[]{\includegraphics[width=0.33\textwidth]{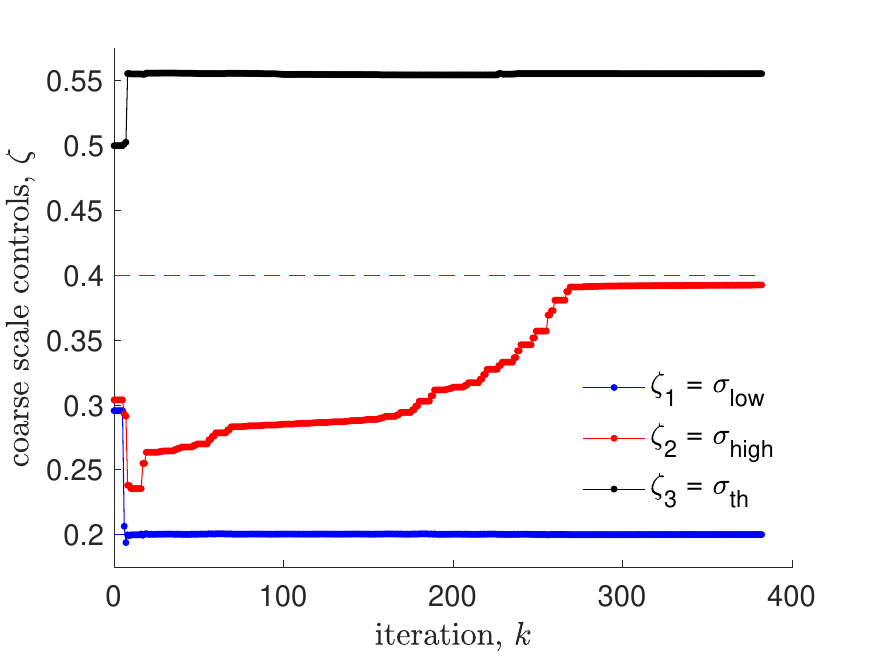}}}
  \mbox{
  \subfigure[]{\includegraphics[width=0.33\textwidth]{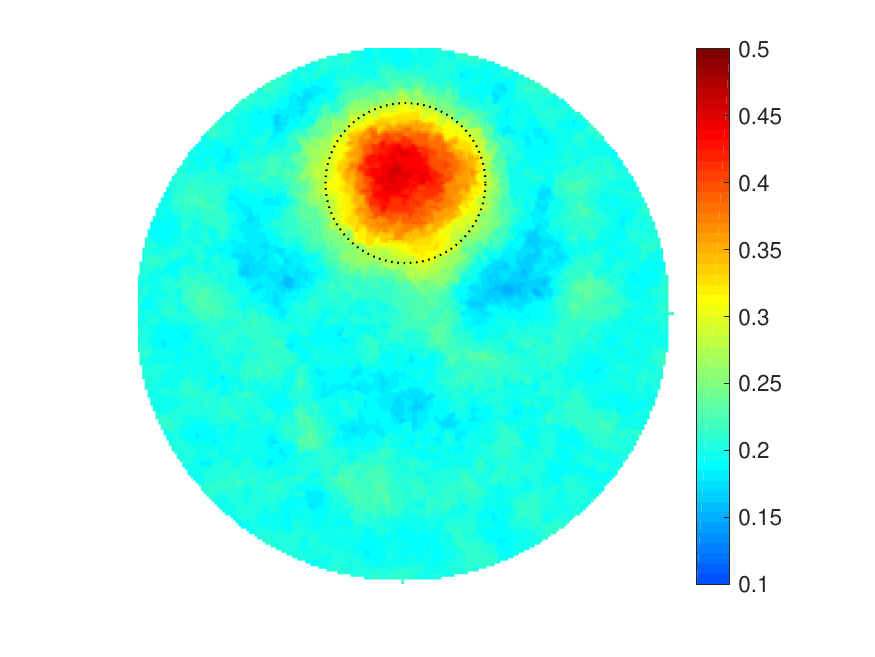}}
  \subfigure[]{\includegraphics[width=0.33\textwidth]{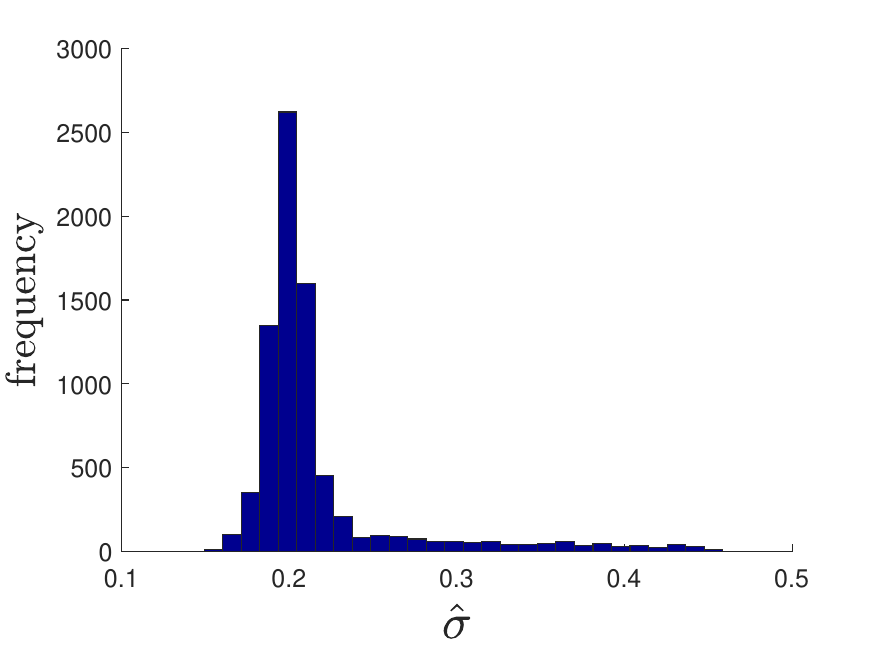}}
  \subfigure[]{\includegraphics[width=0.33\textwidth]{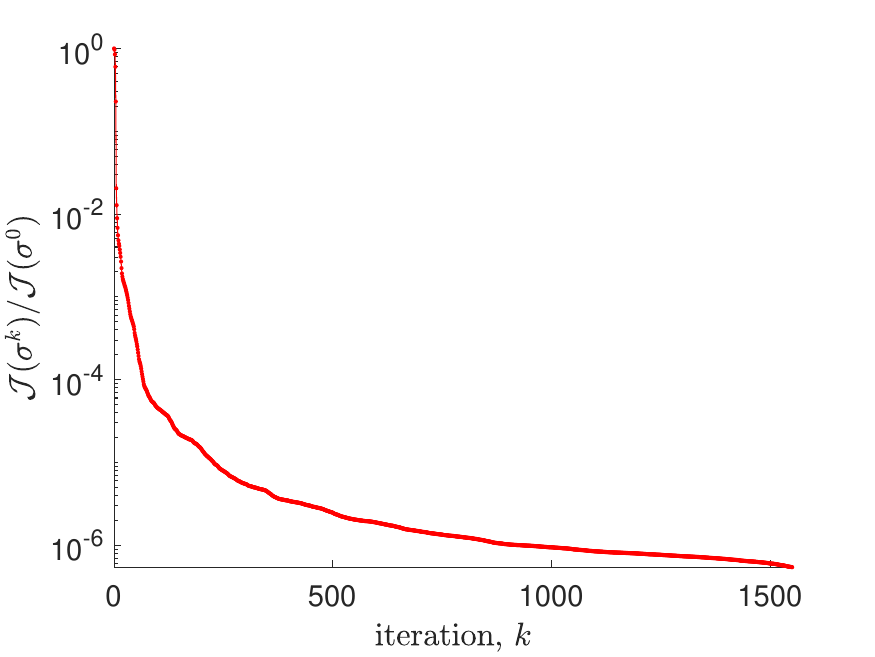}}}
  \end{center}
  \caption{Model \#1 optimization outcomes: (a,b,c)~after applying the multiscale
    framework by Algorithm~\ref{alg:main_opt}, and (d,e,f)~when optimization
    is performed only at a fine scale. Plots in (a,d) show the obtained images with
    added dotted circles to represent the location of the cancer-affected region
    taken from known $\sigma_{true}(x)$ in Figure~\ref{fig:model_9}(a).
    Graphs in (b,f) present normalized cost functional $\mJ(\sigma^k)/\mJ(\sigma^0)$ as
    a function of iteration count $k$ evaluated at fine (red dots) and coarse
    (blue dots) scales. Pink dots in (b) show values $\alpha_{c \rightarrow f}$
    as solutions for problem \eqref{eq:proj_CtoF_rlx}. Changes in the coarse
    scale controls $\zeta^k = [\sigma^k_{low} \ \sigma^k_{high} \ \sigma^k_{th}]$
    are shown in (c) with $\sigma_c = 0.4$ (red dashed line) and $\sigma_h = 0.2$
    (blue dashed line). (e) A histogram constructed for the fine scale solution image (d).}
  \label{fig:model_9_outcome}
\end{figure}

One of the best results in Figure~\ref{fig:model_9_calibr} shown by stars
in fact is obtained with this parameter schedule $\mS^*$.
Figure~\ref{fig:model_9_outcome}(a) shows the outcome of this reconstruction
with a dotted circle representing the location of the cancer-affected region taken
from known $\sigma_{true}(x)$, see Figure~\ref{fig:model_9}(a). The location
of this region is captured accurately. In addition, as seen in
Figure~\ref{fig:model_9_outcome}(c), reconstructed values of both coarse scale
controls $\hat \zeta_1 = \hat \sigma_{low} = 0.2002$ and
$\hat \zeta_2 = \hat \sigma_{high} = 0.3926$ are also very good, although
their rates of convergence differ significantly. The main reason for quick
reconstruction of low conductivity is that it uses superior sensitivity from
gradients computed accurately in close proximity to boundary electrodes.
On another hand, high conductivity regions are usually located deeper in the
interior and span areas in total relatively smaller than ``healthy'' tissues.

Figure~\ref{fig:model_9_outcome}(b) presents normalized cost functional
$\mJ(\sigma^k)/\mJ(\sigma^0)$ as a function of iteration count $k$ evaluated
at both fine (red dots) and coarse (blue dots) scales. Pink dots show
values $\alpha_{c \rightarrow f}$ as solutions for problem \eqref{eq:proj_CtoF_rlx}.
Further analysis of changes in $\alpha_{c \rightarrow f}(k)$ aligned with the
tracked behavior of $\mJ(\sigma^k)$ may suggest more development for termination
conditions to provide even better performance.

Figures~\ref{fig:model_9_outcome}(d,e,f) show also the results of performing
optimization for the same model~\#1 using only fine scale. Following the
discussion in Section~\ref{sec:OPTframe_general} and as seen in
Figure~\ref{fig:model_9_outcome}(e), the second mode corresponding to
$\sigma_h = 0.4$ is not visible in the solution histogram. As a consequence,
a boundary between high and low conductivity regions in domain $\Omega$ is
hardly identifiable by a simple analysis of the solution image and the structure
of its histogram. To add more, using fine scale alone is computationally less
effective as requires more than 1500 iterations to terminate with the same
condition, namely $\epsilon_f = 10^{-10}$. To elaborate more on computational
time, all approaches in this paper use on average 10-12 cost functional evaluations
per iteration for choosing optimal step size in iterative procedures and checking
termination conditions.

Finally, we conclude here that our multiscale computational framework,
when properly calibrated, is able to provide binary images consistent with
the obtained measurements with significant reduction in computational time.

\subsection{Validation with Complicated Models}
\label{sec:model_2_3_4}

We now present results obtained using our new multiscale optimization framework
applied to models with an increased level of complexity. The added complications
are the number of cancer-affected regions (more than one) and the variations in
the size of those regions. Although additional calibration for optimization parameters
might be seen as useful to obtain better results, we use the same parameter schedule
$\mS^*$ in \eqref{eq:opt_schedule} obtained using our benchmark model \#1 as
described in Section~\ref{sec:model_1}.

\begin{figure}[!htb]
  \begin{center}
  \mbox{
  \subfigure[]{\includegraphics[width=0.33\textwidth]{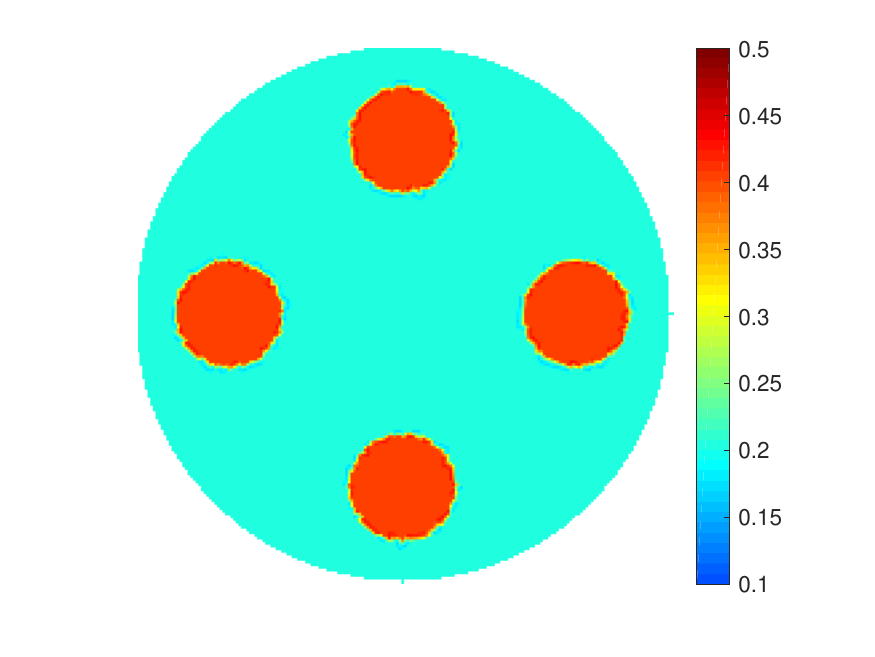}}
  \subfigure[]{\includegraphics[width=0.33\textwidth]{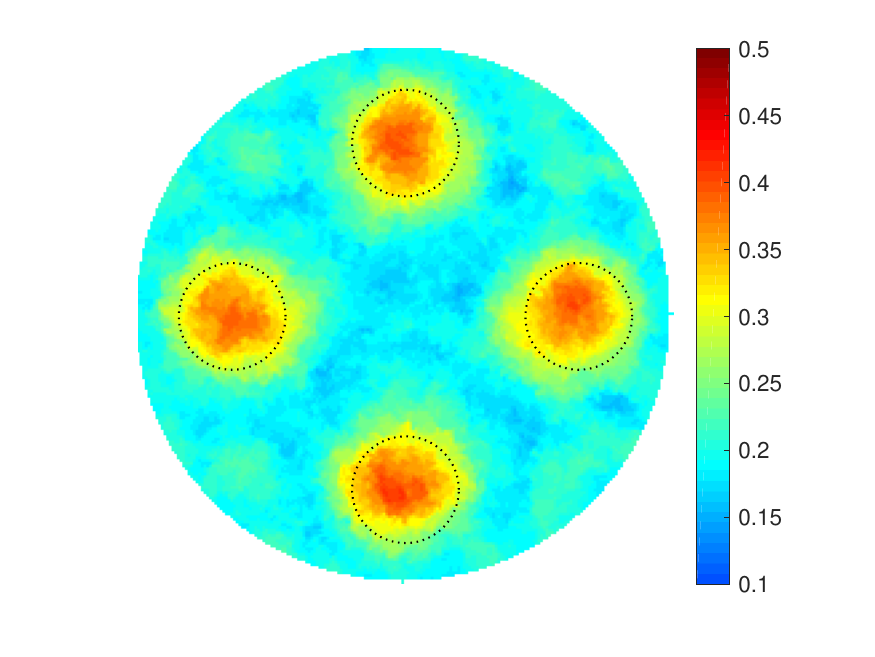}}
  \subfigure[]{\includegraphics[width=0.33\textwidth]{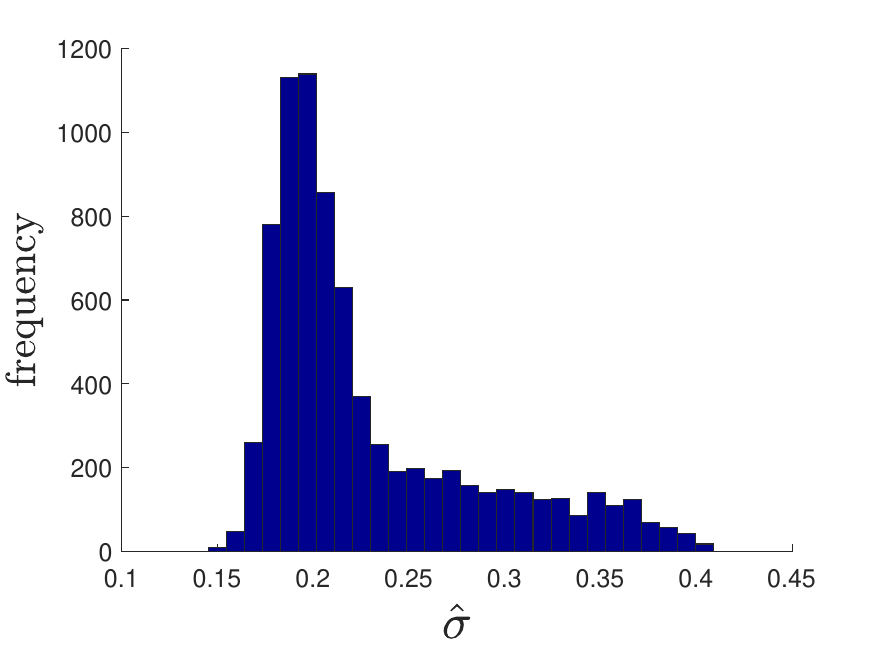}}}
  \mbox{
  \subfigure[]{\includegraphics[width=0.33\textwidth]{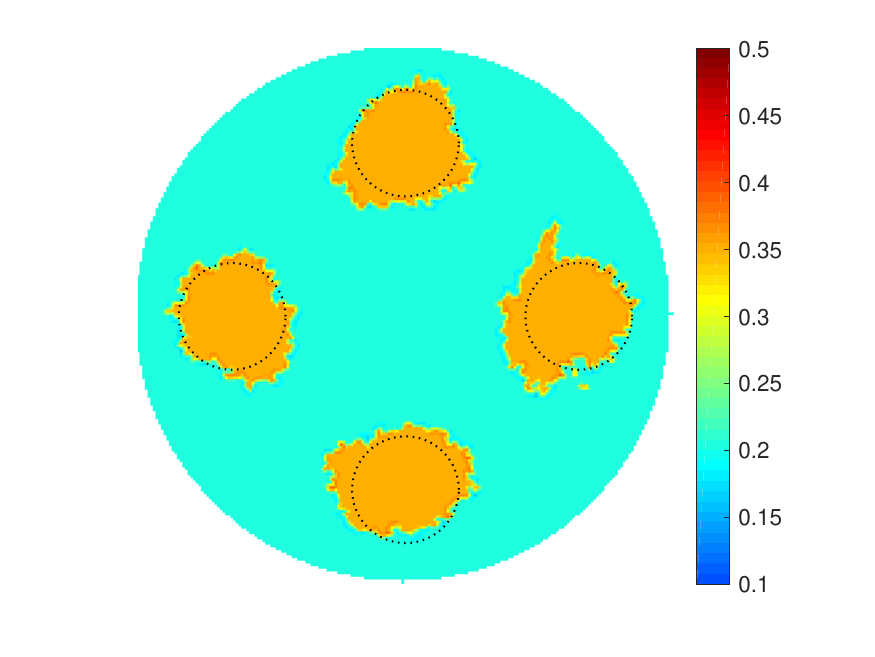}}
  \subfigure[]{\includegraphics[width=0.33\textwidth]{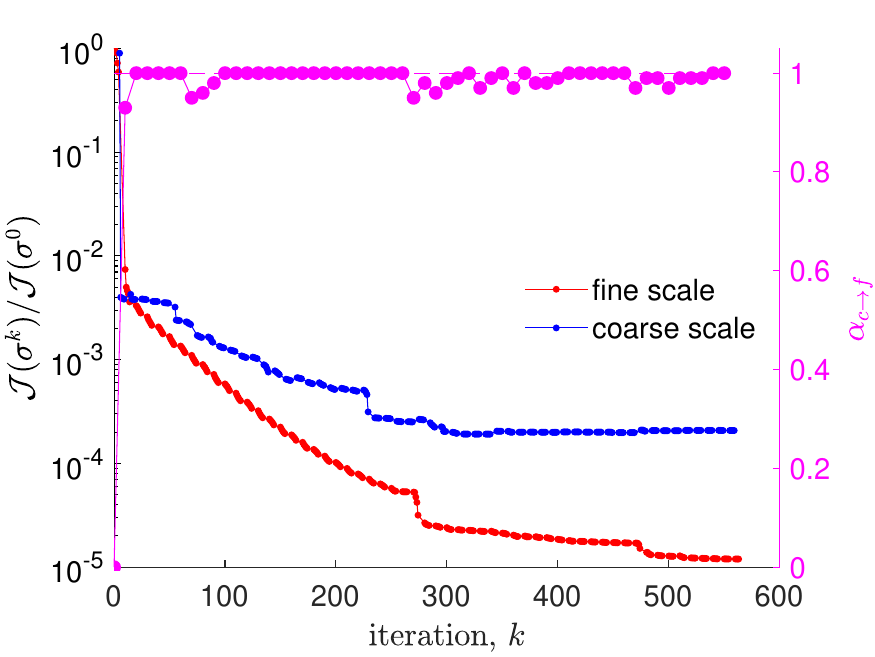}}
  \subfigure[]{\includegraphics[width=0.33\textwidth]{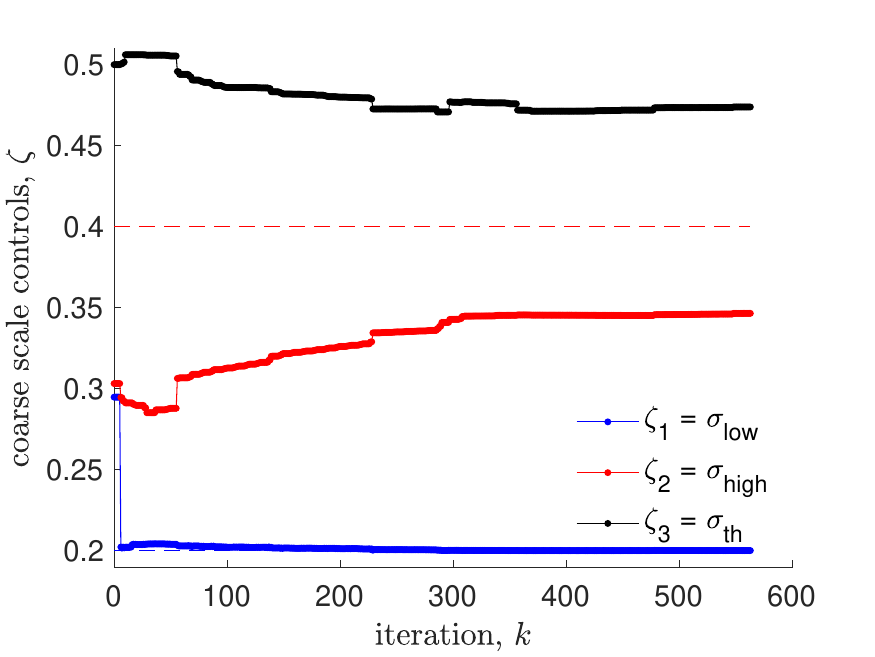}}}
  \end{center}
  \caption{Model \#2. (a)~True electrical conductivity $\sigma_{true}(x)$.
    Outcomes for optimization: (b,c)~when performed only at a fine scale, and
    (d,e,f)~after applying the multiscale framework by Algorithm~\ref{alg:main_opt}.
    Plots in (b,d) show the obtained images with added dotted circles to represent the
    location of four cancer-affected regions taken from known $\sigma_{true}(x)$.
    The fine scale solution histogram is presented in (c). Graph in (e) shows normalized
    cost functional $\mJ(\sigma^k)/\mJ(\sigma^0)$ as a function of iteration count
    $k$ evaluated at fine (red dots) and coarse (blue dots) scales. Pink dots show
    values $\alpha_{c \rightarrow f}$ as solutions for problem
    \eqref{eq:proj_CtoF_rlx}. Changes in the coarse scale controls $\zeta^k =
    [\sigma^k_{low} \ \sigma^k_{high} \ \sigma^k_{th}]$ are shown in (f) with
    $\sigma_c = 0.4$ (red dashed line) and $\sigma_h = 0.2$ (blue dashed line).}
  \label{fig:model_12_outcome}
\end{figure}

The true electrical conductivity $\sigma_{true}(x)$ for our model \#2
containing four same size circular-shaped cancer-affected regions is shown in
Figure~\ref{fig:model_12_outcome}(a). First, we ran a fine scale only optimization
for this model which also required more than 1500 iterations to terminate with
$\epsilon_f = 10^{-10}$. As confirmed by the solution image and the associated
histogram seen in Figures~\ref{fig:model_12_outcome}(b,c), we arrived at the
same conclusion as for model \#1. Even though the high and low conductivity
regions are visualized, clear boundaries between them are hardly identifiable.
On the other hand, as shown in Figure~\ref{fig:model_12_outcome}(d) the binary
image obtained roughly 3 times faster using multiscale optimization locates all
four regions accurately by showing their clear boundaries. Pink and black curves
in Figures~\ref{fig:model_12_outcome}(e,f) show continuous interaction between
scales proving the sensitivity of solutions obtained at one scale to changes
gained at another one. Relaxation parameter $\alpha_{c \rightarrow f}$ (in pink)
different from 1 identifies weighted projections of the coarse scale solutions onto
the fine scale performed with weight $1-\alpha_{c \rightarrow f}$. Updated fine
scale solutions then are used for constructing fine--to--coarse partitions $\mM$ in
\eqref{eq:upscale_map} and \eqref{eq:sigma_coarse} by changing values of the
coarse scale threshold control $\zeta_3 = \sigma_{th}$ (in black). We also
acknowledge the error in recovering high conductivity part $\hat \zeta_2 =
\hat \sigma_{high} = 0.3466$ due to the high non-linearity of the inverse EIT
problem and non-uniqueness of its solution in general, and due to
the presence of ``equivalent'' noise in measurements in particular.

\begin{figure}[!htb]
  \begin{center}
  \mbox{
  \subfigure[]{\includegraphics[width=0.33\textwidth]{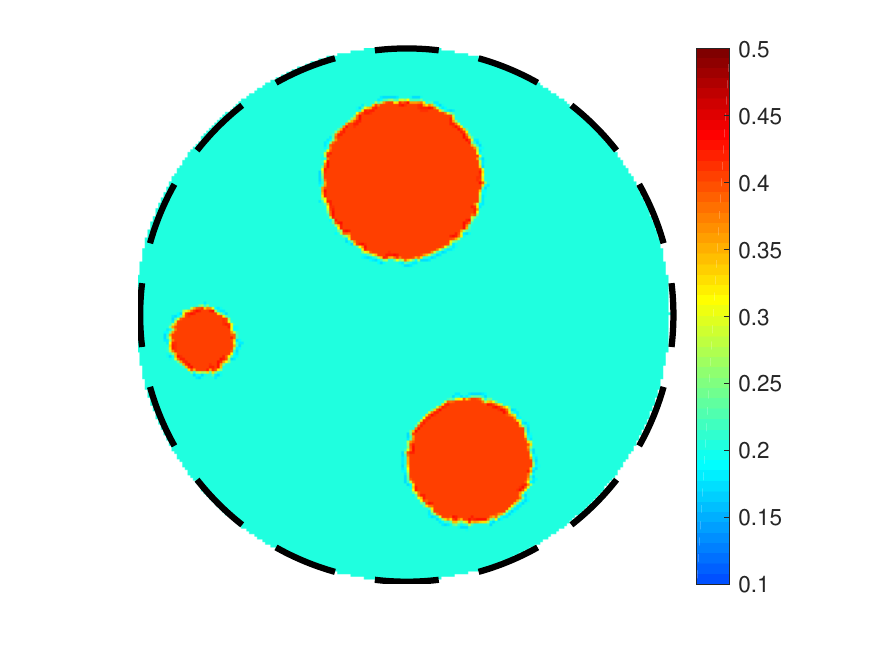}}
  \subfigure[]{\includegraphics[width=0.33\textwidth]{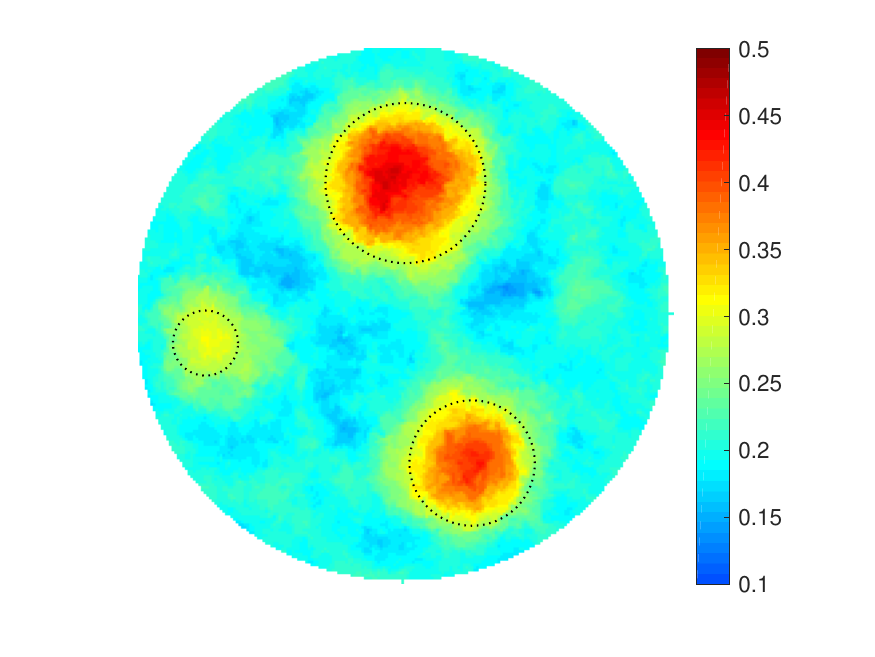}}
  \subfigure[]{\includegraphics[width=0.33\textwidth]{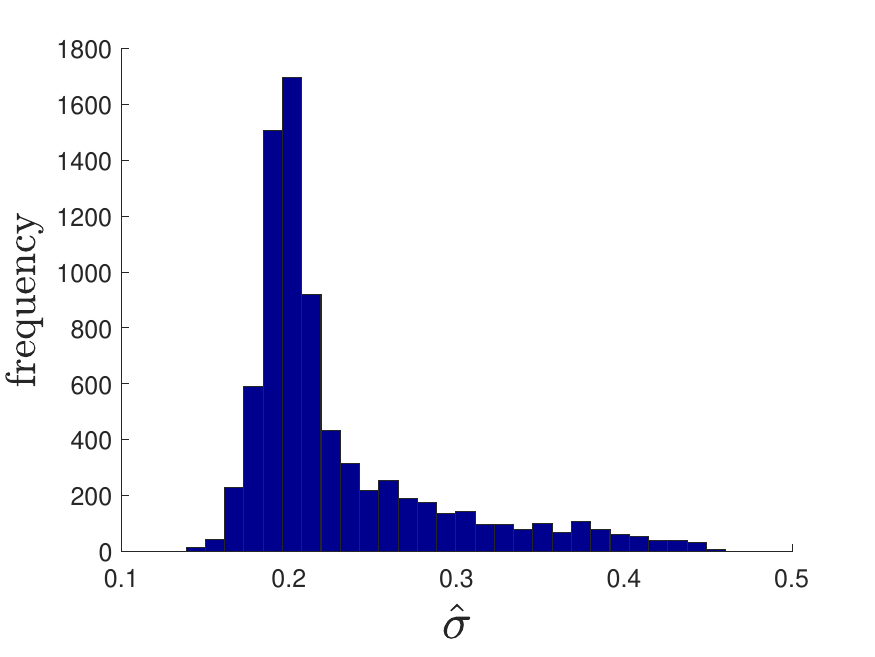}}}
  \mbox{
  \subfigure[]{\includegraphics[width=0.33\textwidth]{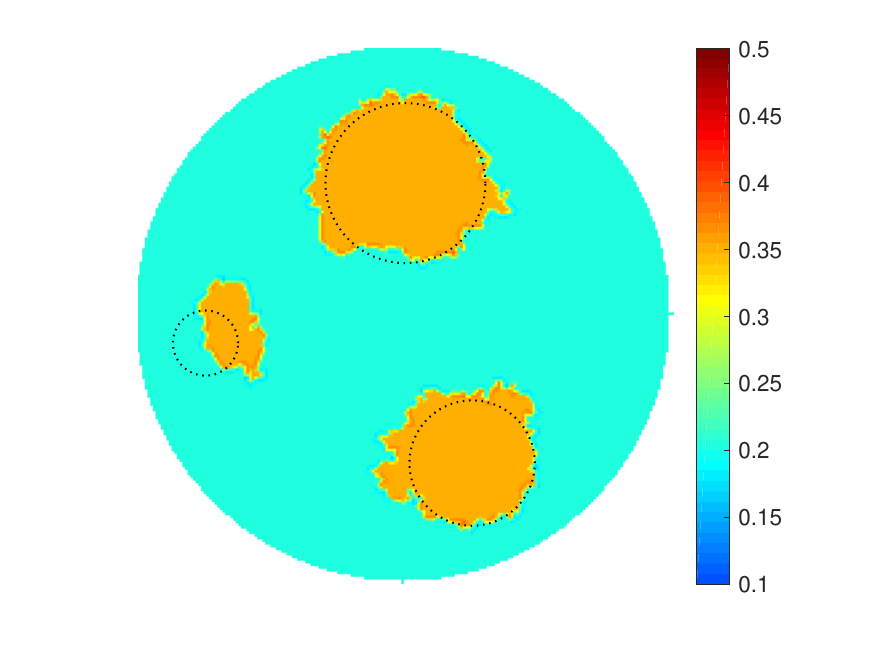}}
  \subfigure[]{\includegraphics[width=0.33\textwidth]{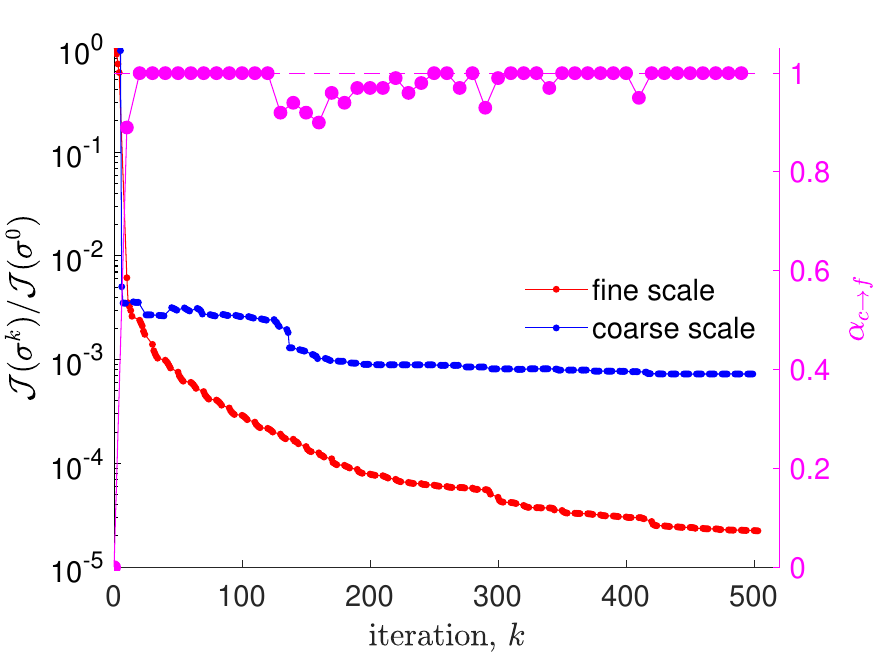}}
  \subfigure[]{\includegraphics[width=0.33\textwidth]{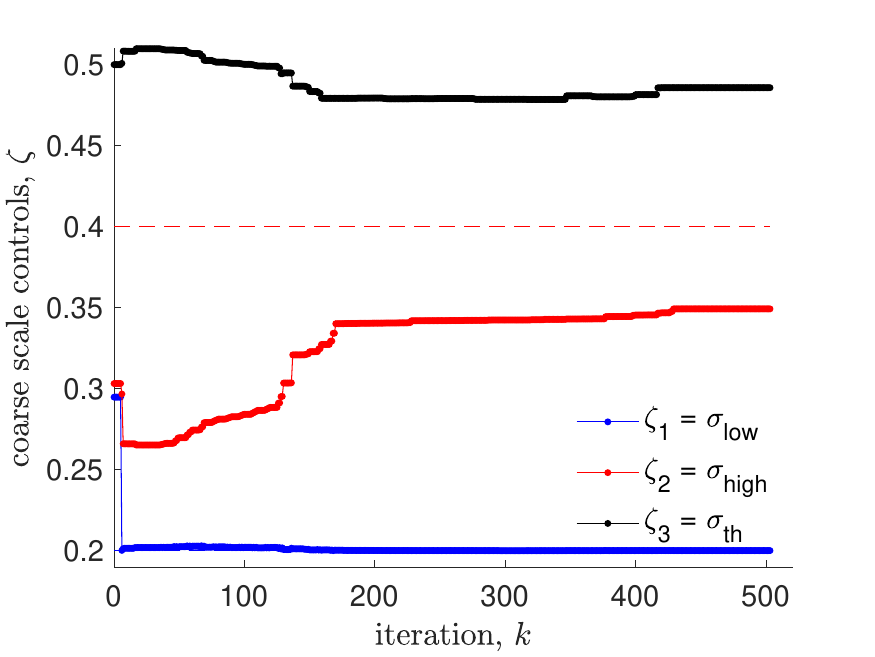}}}
  \end{center}
  \caption{Model \#3. (a)~True electrical conductivity $\sigma_{true}(x)$ with
    added structure of boundary electrodes (in black).
    Outcomes for optimization: (b,c)~when performed only at a fine scale, and
    (d,e,f)~after applying the multiscale framework by Algorithm~\ref{alg:main_opt}.
    Plots in (b,d) show the obtained images with added dotted circles to represent the
    location of three cancer-affected regions taken from known $\sigma_{true}(x)$.
    The fine scale solution histogram is presented in~(c). Graph in (e) shows normalized
    cost functional $\mJ(\sigma^k)/\mJ(\sigma^0)$ as a function of iteration count
    $k$ evaluated at fine (red dots) and coarse (blue dots) scales. Pink dots show
    values $\alpha_{c \rightarrow f}$ as solutions for problem
    \eqref{eq:proj_CtoF_rlx}. Changes in the coarse scale controls $\zeta^k =
    [\sigma^k_{low} \ \sigma^k_{high} \ \sigma^k_{th}]$ are shown in (f) with
    $\sigma_c = 0.4$ (red dashed line) and $\sigma_h = 0.2$ (blue dashed line).}
  \label{fig:model_8_outcome}
\end{figure}

Our next model \#3 contains three circular-shaped cancer-affected regions
of various sizes. Its electrical conductivity $\sigma_{true}(x)$ is shown
in Figure~\ref{fig:model_8_outcome}(a). As in the previous case, we run a fine
scale only optimization terminated with $\epsilon_f = 10^{-10}$ which gives
a solution image and the associated histogram as shown in
Figures~\ref{fig:model_8_outcome}(b,c). Compared with \#2, this model is
considered harder as it contains a small spot at the left whose dimension
is comparable with the size of the boundary electrodes added in black to
Figure~\ref{fig:model_8_outcome}(a). The image obtained at the fine scale,
see Figure~\ref{fig:model_8_outcome}(b), confirms this complexity as,
unlike the two bigger spots, the smallest one lost its color development and,
consequently, could be missed. Further analysis of
Figures~\ref{fig:model_8_outcome}(b,c) suggests that clear boundaries between
the high and low conductivity regions are hardly identifiable.

Similarly to model \#2, Figure~\ref{fig:model_8_outcome}(d) exhibits results of
applying multiscale optimization Algorithm~\ref{alg:main_opt} as a clear image of
model \#3 with a nice binary resolution enabled to locate three cancer-affected
regions. Continuous information exchange between fine and coarse scales due to
solution projections is also observed in Figures~\ref{fig:model_8_outcome}(e,f).
Here we should comment that the deviated location of the smallest region and
the error in recovering high conductivity part of the coarse scale control
$\zeta$, $\hat \zeta_2 = \hat \sigma_{high} = 0.3493$, could be also explained
by the high non-linearity of the inverse EIT problem and the presence of the
``equivalent'' noise. To add more, various size and using $\sigma_{th}$ as a single
control for all three regions may significantly contribute to amplify these errors.

\begin{figure}[!htb]
  \begin{center}
  \mbox{
  \subfigure[]{\includegraphics[width=0.33\textwidth]{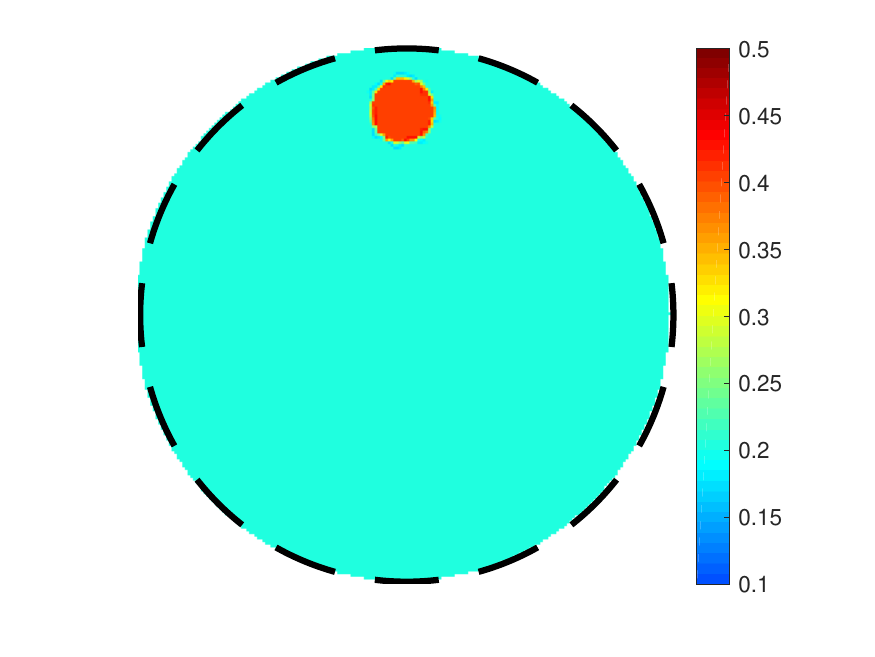}}
  \subfigure[]{\includegraphics[width=0.33\textwidth]{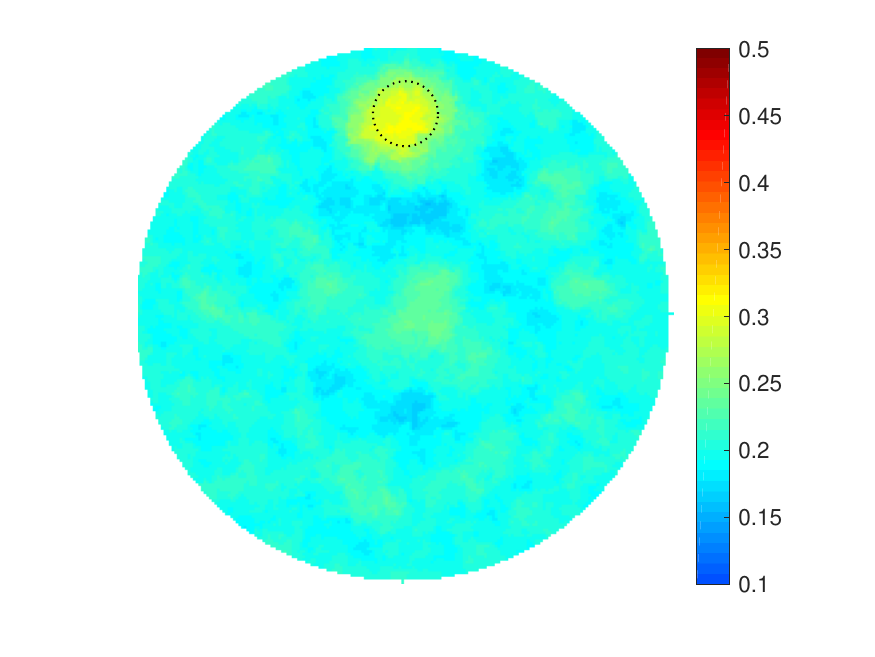}}
  \subfigure[]{\includegraphics[width=0.33\textwidth]{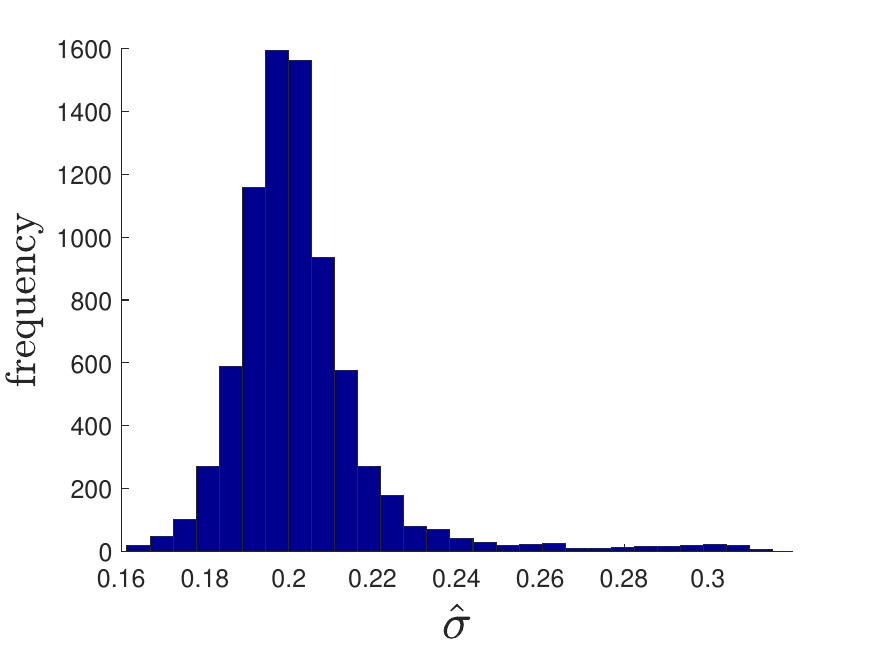}}}
  \mbox{
  \subfigure[]{\includegraphics[width=0.33\textwidth]{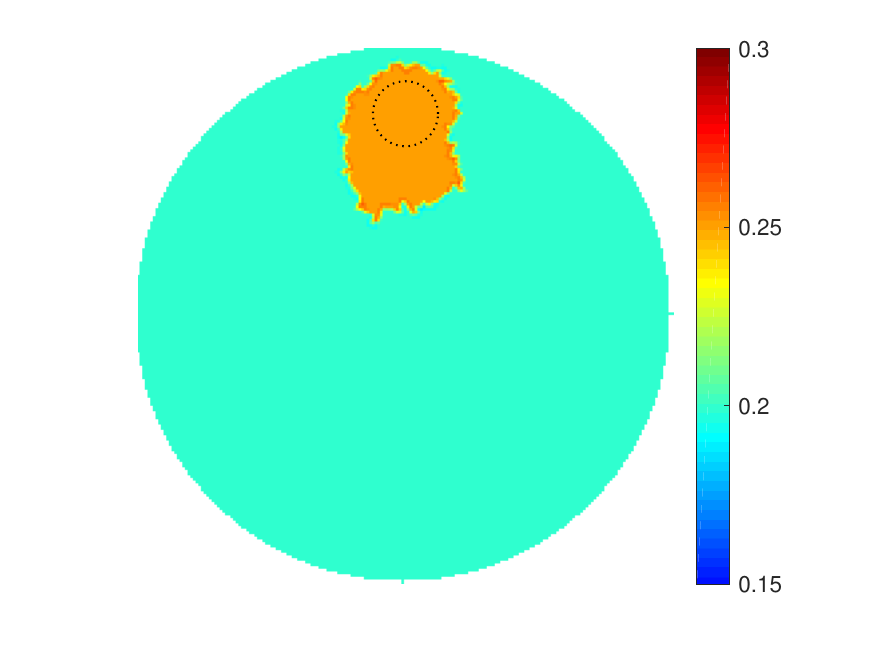}}
  \subfigure[]{\includegraphics[width=0.33\textwidth]{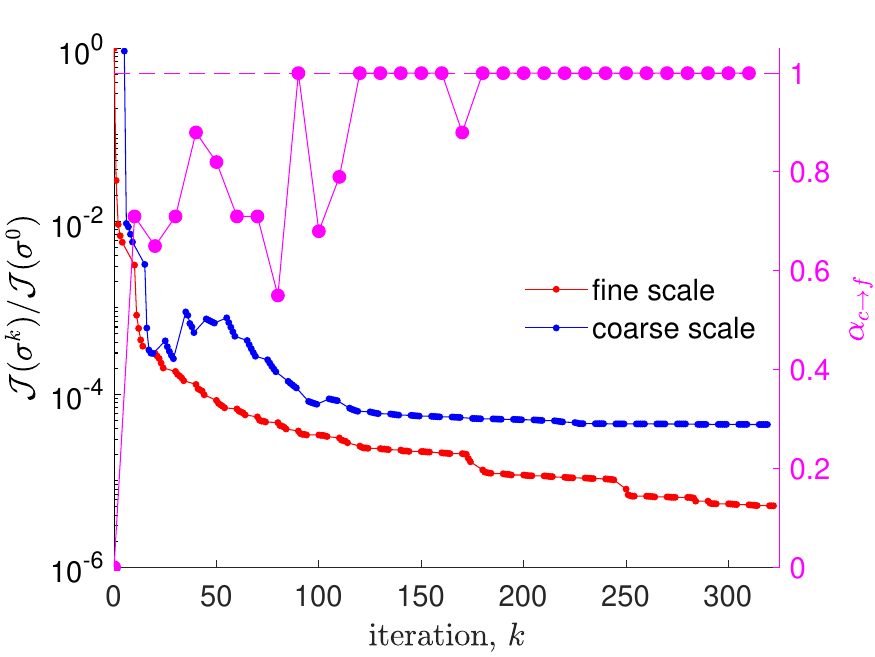}}
  \subfigure[]{\includegraphics[width=0.33\textwidth]{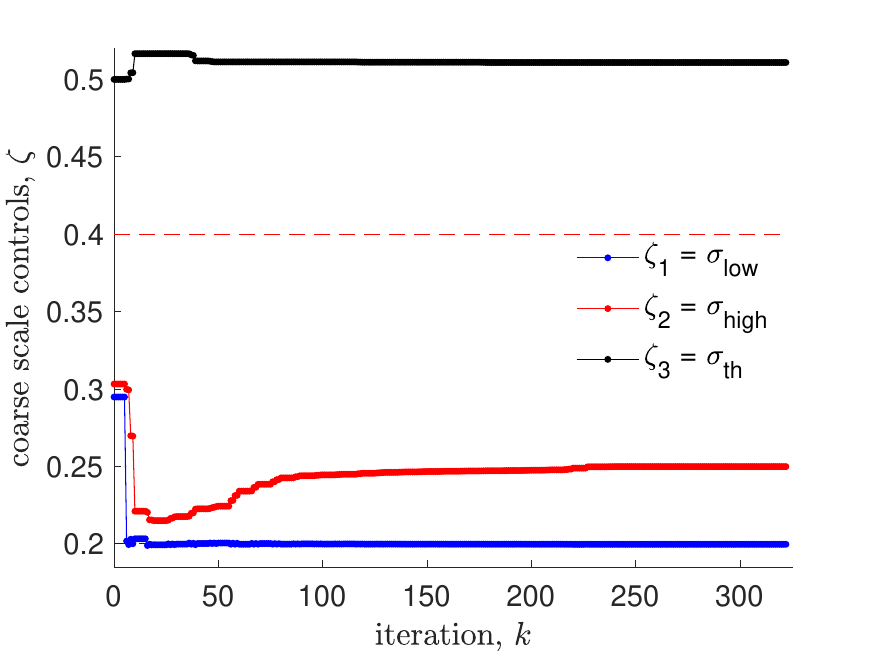}}}
  \end{center}
  \caption{Model \#4. (a)~True electrical conductivity $\sigma_{true}(x)$ with
    added structure of boundary electrodes (in black).
    Outcomes for optimization: (b,c)~when performed only at a fine scale, and
    (d,e,f)~after applying the multiscale framework by Algorithm~\ref{alg:main_opt}.
    Plots in (b,d) show the obtained images with added dotted circles to represent the
    location of one small cancer-affected region taken from known $\sigma_{true}(x)$.
    Graph in (e) shows normalized
    cost functional $\mJ(\sigma^k)/\mJ(\sigma^0)$ as a function of iteration count
    $k$ evaluated at fine (red dots) and coarse (blue dots) scales. Pink dots show
    values $\alpha_{c \rightarrow f}$ as solutions for problem
    \eqref{eq:proj_CtoF_rlx}. Changes in the coarse scale controls $\zeta^k =
    [\sigma^k_{low} \ \sigma^k_{high} \ \sigma^k_{th}]$ are shown in (f) with
    $\sigma_c = 0.4$ (red dashed line) and $\sigma_h = 0.2$ (blue dashed line).}
  \label{fig:model_13_outcome}
\end{figure}

Our last model \#4 is the hardest one created to mimic using the EIT techniques
in medical practice for recognizing cancer at the early stages. The electrical
conductivity $\sigma_{true}(x)$ is shown in Figure~\ref{fig:model_13_outcome}(a).
This model contains only one circular-shaped cancer-affected region of the same
size as the smallest region in model \#3. The known complication comes from the
fact that the order of difference in measurements generated by this model and
``healthy tissue'' ($\sigma(x) = \sigma_h, \ \forall x \in \Omega$) is very
close to the order of noise that appeared naturally in provided data. As for previous
models, we compare the results obtained after applying fine scale only optimization,
see Figures~\ref{fig:model_13_outcome}(b,c), and multiscale optimization
Algorithm~\ref{alg:main_opt}, shown in Figures~\ref{fig:model_13_outcome}(d,e,f),
both terminated with $\epsilon_f = 10^{-10}$.

After comparing fine scale and multiscale optimization images, it is true to
say that the latter could provide more assistance in concluding on possible
abnormal changes in tissues and navigating the surgeons. On the other hand,
the fine scale image may be misleading as some (yellowish) regions may be
deceptively interpreted as being cancerous. The multiscale optimization,
however, allows keeping its images devoid of this problem and, as such,
shows high potential in minimizing possibilities for false positive screening
and improving the overall quality of the EIT-based procedures.
Similarly to our previous models, the same conclusion arrives after observing
continuous information exchange between fine and coarse scales seen in
Figures~\ref{fig:model_13_outcome}(e,f) which results in creating an image
with clear binary resolution ready to locate the cancer-affected region.
We also note that its size appears larger than in the true model and the error
in recovering the coarse scale control $\hat \zeta_2 = \hat \sigma_{high}
= 0.2499$ is much bigger than that seen in the previous models. However,
we expect these results may be further improved by applying additional calibration
to the (potentially extended) list of chosen parameters, as well as by applying
more advanced minimization techniques while performing optimization at both
fine and coarse scales.

\section{Concluding Remarks}
\label{sec:remarks}

In this work, we presented an efficient computational approach for optimal
reconstructing the physical properties of the media characterized by distributions
close to binary. In particular, this approach could be useful in various applications
in biomedical sciences to operate with physical models supplied with some,
possibly incomplete and noisy, measurements. This claim is supported by a
simple fact that the proposed multiscale algorithm is given in a very broad context
and could be easily applied to other models by changing PDE systems in the formulation
of forward and adjoint problems. The proposed computational framework
includes an array of gradient-based multiscale optimization techniques supplied
with multilevel control space reduction over both fine and coarse scales used
interchangeably. Quality and computational efficiency of the obtained results
are ensured by developing a methodology for establishing an effective
``communication'' between scales by projecting the solutions from one scale
onto another and accumulating optimally progress obtained at both scales.
Such multiscale optimization paired with multilevel control space
reduction allows using computational advantages seen at both scales and to
mitigate their negative impacts.

We investigated the performance of our complete computational framework
in applications to the 2D inverse problem of cancer detection by the
Electrical Impedance Tomography technique. Our first benchmark model mimics
a simple biological tissue case with confirmed presence of one circular-shaped
area affected by cancer. The proposed procedure for calibrating certain parameters
was applied to this model to ensure the enhanced performance of our optimization
framework. We also presented results obtained by applying the calibrated framework
to multiple models of an increased level of complexity, namely with three and for
cancer-affected regions of various sizes. For every model, we obtained clear images
with a nice binary resolution enabled to locate all cancer-affected regions.
In addition, our multiscale optimization framework proved its high efficiency
by completing computations 3 times faster than in cases when only fine scale
was in use.

We also check the applicability of our framework in applications to procedures
for cancer recognition at the early stages by a model containing one tiny cancerous
spot with a diameter comparable with the size of the boundary electrodes. Despite
the errors in recovering the true shape and the values of electrical conductivity,
we conclude that obtained images of that quality will provide valuable assistance
in recognizing possible abnormal changes in tissues and further navigating medical
professionals with their decisions. We conclude that the properly calibrated
multiscale optimization framework is able to provide binary images consistent with
the provided measurements using significantly reduced computational time. In general,
we see a high potential of the proposed computational framework in minimizing
possibilities for false positive screening and improving the overall quality of
the EIT-based procedures.

There are many ways in which our multiscale optimization algorithm can be tested
and extended. We provided an example of a simple calibration procedure, but we
expect the performance may be further improved by extending the list of calibrated
parameters and applying more advanced minimization techniques to perform local and
global searches while performing optimization at both fine and coarse scales.
Given that we used data provided by a specific electrode configuration, it will
be of interest to apply a further analysis of the measurement structure, for example
considering a 32-electrode scheme and improving sensitivity by optimizing the
structure of available data. Also, as many modern EIT systems feature pair-wise
voltage patterns, we will be interested in testing the performance of our new
method in applications to such systems. We also plan to investigate the use of
flexible schemes for switching between scales including new approaches for projecting
solutions. The impact of the noise present in measurements should be also
systematically analyzed. Also of interest is the extension of our multiscale
optimization approach to include possibility of using various PCA sample
structures, multiple coarse scale controls associated with different spatial
regions, and applicability to bimodal distributions.

Finally, it is important to test our new approach in various applications to
real data and different types of cancerous tissues, as this would certainly
suggest areas in which further developments may be required. For example,
some applications of EIT may require modeling conductivity with more than
one different values to characterize ``healthy'' and ''cancerous'' tissues.
We expect our new approach could be easily adapted to this and even more
complicated settings, e.g.~considering electrical conductivity to be
considered fully anisotropic, seen in reality. Despite the fact
that this approach was initially tested with synthetic EIT-related problems,
we believe that this methodology could be easily applied to a broad range of
problems in biomedical sciences, also in physics, geology, chemistry, etc.

\bibliographystyle{spmpsci}
\bibliography{biblio_Bukshtynov,biblio_EIT,biblio_OPT}

\begin{thebibliography}{10}
\providecommand{\url}[1]{{#1}}
\providecommand{\urlprefix}{URL }
\expandafter\ifx\csname urlstyle\endcsname\relax
  \providecommand{\doi}[1]{DOI~\discretionary{}{}{}#1}\else
  \providecommand{\doi}{DOI~\discretionary{}{}{}\begingroup
  \urlstyle{rm}\Url}\fi

\bibitem{Abascal2011}
Abascal, J.F.P.J., Lionheart, W.R.B., Arridge, S.R., Schweiger, M., Atkinson,
  D., Holder, D.S.: Electrical impedance tomography in anisotropic media with
  known eigenvectors.
\newblock Inverse Problems \textbf{27}(6), 1--17 (2011)

\bibitem{AbdullaBukshtynovSeif}
Abdulla, U.G., Bukshtynov, V., Seif, S.: Breast cancer detection through
  electrical impedance tomography and optimal control theory: Theoretical and
  computational analysis.
\newblock \url{arXiv:1809.05936}

\bibitem{Adler2008}
Adler, A., Arnold, J., Bayford, R., Borsic, A., Brown, B., Dixon, P., Faes,
  T.J., Frerichs, I., Gagnon, H., G{\"a}rber, Y., Grychtol, B., Hahn, G.,
  Lionheart, W., Malik, A., Stocks, J., Tizzard, A., Weiler, N., Wolf, G.:
  {GREIT:} towards a consensus {EIT} algorithm for lung images.
\newblock In: 9th {EIT} conference 2008, 16-18 June 2008, Dartmouth, New
  Hampshire. Citeseer (2008)

\bibitem{Adler2015}
Adler, A., Gaburro, R., Lionheart, W.: Handbook of Mathematical Methods in
  Imaging, chap. Electrical Impedance Tomography, pp. 701--762.
\newblock Springer New York, New York, NY (2015)

\bibitem{Alber2019}
Alber, M., Tepole, A.B., Cannon, W.R., De, S., Dura-Bernal, S., Garikipati, K.,
  Karniadakis, G., Lytton, W.W., Perdikaris, P., Petzold, L., Kuhl, E.:
  Integrating machine learning and multiscale modeling -- perspectives,
  challenges, and opportunities in the biological, biomedical, and behavioral
  sciences.
\newblock Digital Medicine \textbf{2}(115) (2019)

\bibitem{Bera2018}
Bera, T.K.: Applications of electrical impedance tomography ({EIT}): A short
  review.
\newblock {IOP} Conference Series: Materials Science and Engineering
  \textbf{331}, 012,004 (2018)

\bibitem{Berger1977}
Berger, M.S.: Nonlinearity and Functional Analysis.
\newblock Acad.~Press, New York (1977)

\bibitem{Borcea2002}
Borcea, L.: Electrical impedance tomography.
\newblock Inverse Problems \textbf{18}, 99--136 (2002)

\bibitem{Brown2003}
Brown, B.: Electrical impedance tomography {(EIT)}: A review.
\newblock Journal of Medical Engineering and Technology \textbf{27}(3), 97--108
  (2003)

\bibitem{Bukshtynov15}
Bukshtynov, V., Volkov, O., Durlofsky, L., Aziz, K.: Comprehensive framework
  for gradient-based optimization in closed-loop reservoir management.
\newblock Computational Geosciences \textbf{19}(4), 877--897 (2015)

\bibitem{Calderon1980}
Calderon, A.P.: On an inverse boundary value problem.
\newblock In: Seminar on Numerical Analysis and Its Applications to Continuum
  Physics, pp. 65--73. Soc.~Brasileira de Mathematica, Rio de Janeiro (1980)

\bibitem{ChenChengLinWang2009}
Chen, W., Cheng, J., Lin, J., Wang, L.: A level set method to reconstruct the
  discontinuity of the conductivity in {EIT}.
\newblock Science in China Series {A}: Mathematics \textbf{52}, 29--44 (2009)

\bibitem{Cheney1999}
Cheney, M., Isaacson, D., Newell, J.: Electrical impedance tomography.
\newblock SIAM Review \textbf{41}(1), 85--101 (1999)

\bibitem{Cheng1989}
Cheng, K.S., Isaacson, D., Newell, J., Gisser, D.G.: Electrode models for
  electric current computed tomography.
\newblock {IEEE} Transactions on Biomedical Engineering \textbf{36}(9),
  918--924 (1989)

\bibitem{Clancy2016}
Clancy, C.E., An, G., Cannon, W.R., Liu, Y., May, E.E., Ortoleva, P., Popel,
  A.S., Sluka, J.P., Su, J., Vicini, P., Zhou, X., Eckmann, D.M.: Multiscale
  modeling in the clinic: {D}rug design and development.
\newblock Annals of Biomedical Engineering \textbf{44}(9), 2591--2610 (2016)

\bibitem{Cominelli2007}
Cominelli, A., Ferdinandi, F., De~Montleau, P., Rossi, R.: Using gradients to
  refine parameterization in field-case history-matching projects.
\newblock SPE Reservoir Evaluation and Engineering \textbf{10}(3), 233--240
  (2007)

\bibitem{UMFPACK}
Davis, T.A.: Algorithm 832: {UMFPACK V4.3} -- an unsymmetric-pattern
  multifrontal method.
\newblock ACM Transactions on Mathematical Software (TOMS) \textbf{30}(2),
  196--199 (2004)

\bibitem{GibouFedkiwOsher2018}
Gibou, F., Fedkiw, R., Osher, S.: A review of level-set methods and some recent
  applications.
\newblock Journal of Computational Physics \textbf{353}, 82--109 (2018)

\bibitem{Grimstad2000}
Grimstad, A.A., Mannseth, T.: Nonlinearity, scale, and sensitivity for
  parameter estimation problems.
\newblock SIAM Journal on Scientific Computing \textbf{21}(6), 2096--2113
  (2000)

\bibitem{Grimstad2003}
Grimstad, A.A., Mannseth, T., N{\ae}vdal, G., Urkedal, H.: Adaptive multiscale
  permeability estimation.
\newblock Computational Geosciences \textbf{7}, 1--25 (2003)

\bibitem{FreeFem2012}
Hecht, F.: New development in {FreeFem++}.
\newblock Journal of Numerical Mathematics \textbf{20}(3-4), 251--265 (2012)

\bibitem{Holder2004}
Holder, D.S.: Electrical Impedance Tomography. Methods, History and
  Applications.
\newblock CRC Press (2004)

\bibitem{Horstemeyer2009}
Horstemeyer, M.F.: Practical Aspects of Computational Chemistry, chap.
  Multiscale Modeling: {A} Review.
\newblock Springer (2009)

\bibitem{Jolliffe2002}
Jolliffe, I.T.: Principal Component Analysis, Second Edition.
\newblock Springer (2002)

\bibitem{JolliffeCadima2016}
Jolliffe, I.T., Cadima, J.: Principal component analysis: a review and recent
  developments.
\newblock Phil. Trans. R. Soc. A. \textbf{374}(2065) (2016)

\bibitem{Lien2005}
Lien, M., Berre, I., Mannseth, T.: Combined adaptive multiscale and level-set
  parameter estimation.
\newblock Multiscale Modeling \& Simulation \textbf{4}(4), 1349--1372 (2005)

\bibitem{Lionheart2004}
Lionheart, W.: {EIT} reconstruction algorithms: Pitfalls, challenges and recent
  developments.
\newblock Physiological Measurement \textbf{25}(1), 125--142 (2004)

\bibitem{Liu2018}
Liu, D., Khambampati, A.K., Du, J.: A parametric level set method for
  electrical impedance tomography.
\newblock {IEEE} Transactions on Medical Imaging \textbf{37}(2), 451--460
  (2018)

\bibitem{Nocedal2006}
Nocedal, J., Wright, S.J.: Numerical Optimization, 2nd edn.
\newblock Springer (2006)

\bibitem{OsherSethian1988}
Osher, S., Sethian, J.: Fronts propagating with curvature-dependent speed:
  Algorithms based on {H}amilton-{J}acobi formulations.
\newblock Journal of Computational Physics \textbf{79}(1), 12--49 (1988)

\bibitem{NumericalRecipes2007}
Press, W.H., Teukolsky, S.A., Vetterling, W.T., Flannery, B.P.: Numerical
  Recipes: {T}he Art of Scientific Computing, 3rd edn.
\newblock Cambridge University Press (2007)

\bibitem{Steinhauser2017}
Steinhauser, M.: Computational Multiscale Modeling of Fluids and Solids:
  {T}heory and Applications, 2nd edn.
\newblock Springer (2017)

\bibitem{Tai2004}
Tai, X.C., Chan, T.: A survey on multiple level set methods with applications
  for identifying piecewise constant functions.
\newblock International Journal of Numerical Analysis and Modeling
  \textbf{1}(1), 25--47 (2004)

\bibitem{Tawhai2009}
Tawhai, M., Bischoff, J., Einstein, D., Erdemir, A., Guess, T., Reinbolt, J.:
  Multiscale modeling in computational biomechanics.
\newblock {IEEE} Engineering in Medicine and Biology Magazine \textbf{28}(3),
  41--49 (2009)

\bibitem{TsaiOsher2003}
Tsai, R., Osher, S.: Level set methods and their applications in image science.
\newblock Communications in Mathematical Sciences \textbf{1}(4), 1--20 (2003)

\bibitem{Uhlmann2009}
Uhlmann, G.: Electrical impedance tomography and {C}alder{\'o}n's problem.
\newblock Inverse Problems \textbf{25}(12), 123,011 (2009)

\bibitem{VolkovBukshtynov18}
Volkov, O., Bukshtynov, V., Durlofsky, L., Aziz, K.: Gradient-based {P}areto
  optimal history matching for noisy data of multiple types.
\newblock Computational Geosciences \textbf{22}(6), 1465--1485 (2018)

\bibitem{Walpole2013}
Walpole, J., Papin, J.A., Peirce, S.M.: Multiscale computational models of
  complex biological systems.
\newblock Annual Review of Biomedical Engineering \textbf{15}, 137--154 (2013)

\bibitem{Wang2020}
Wang, Z., Yue, S., Wang, H., Wang, Y.: Data preprocessing methods for
  electrical impedance tomography: a review.
\newblock Physiological Measurement \textbf{41}(9), 09TR02 (2020)

\bibitem{Weinan2011}
Weinan, E.: Principles of Multiscale Modeling.
\newblock Cambridge University Press (2011)

\bibitem{Zou2003Review}
Zou, Y., Guo, Z.: A review of electrical impedance techniques for breast cancer
  detection.
\newblock Medical Engineering and Physics \textbf{25}(2), 79--90 (2003)

\end{thebibliography}

\end{document}